\documentclass[preprint]{emulateapj}

\pdfoutput=1

\newcommand{\nustar}{{\it NuSTAR}}
\newcommand{\swift}{{\it Swift}}
\newcommand{\chandra}{{\it Chandra}}
\newcommand{\xmm}{{\it XMM-Newton}}
\newcommand{\suzaku}{{\it Suzaku}}
\newcommand{\etal}{{\it et al.}}

\def\arcsec{$\,^{\prime\prime}$}
\def\deg{$^\circ$}
\def\arcmin{$\,^{\prime}$}

\def\simless{\mathbin{\lower 3pt\hbox
   {$\rlap{\raise 5pt\hbox{$\char'074$}}\mathchar"7218$}}} 
\def\simgreat{\mathbin{\lower 3pt\hbox
   {$\rlap{\raise 5pt\hbox{$\char'076$}}\mathchar"7218$}}} 

\slugcomment{Submitted to The Astrophysical Journal.}

\shorttitle{The Nuclear Spectroscopic Telescope Array (NuSTAR) Mission}
\shortauthors{Harrison et al.}

\begin{document}

\title{The {\it Nuclear Spectroscopic Telescope Array (NuSTAR)} High Energy X-ray Mission}


\author{Fiona A. Harrison\altaffilmark{1}, 
William W. Craig\altaffilmark{2,3},  
Finn E. Christensen\altaffilmark{4},
Charles J. Hailey\altaffilmark{5}, 
William W. Zhang\altaffilmark{6}, 
Steven E. Boggs\altaffilmark{3}, 
Daniel Stern\altaffilmark{7}, 
W. Rick Cook\altaffilmark{1},
Karl Forster\altaffilmark{1},
Paolo Giommi\altaffilmark{8,9},
Brian W. Grefenstette\altaffilmark{1},
Yunjin Kim\altaffilmark{7},
Takao Kitaguchi\altaffilmark{10},
Jason E Koglin\altaffilmark{5,11},
Kristin K. Madsen\altaffilmark{1},
Peter H. Mao,\altaffilmark{1},
Hiromasa Miyasaka\altaffilmark{1},
Kaya Mori\altaffilmark{5},
Matteo Perri\altaffilmark{8,9},
Michael J. Pivovaroff\altaffilmark{2},
Simonetta Puccetti\altaffilmark{8,9},
Vikram R.  Rana\altaffilmark{1},
Niels J. Westergaard\altaffilmark{4},
Jason Willis\altaffilmark{7},
Andreas Zoglauer\altaffilmark{3},
Hongjun An\altaffilmark{12}, 
Matteo Bachetti\altaffilmark{13},
Nicolas~M. Barri{\`e}re\altaffilmark{3}, 
Eric C. Bellm \altaffilmark{1}, 
Varun Bhalerao\altaffilmark{1,14},
Nicolai F. Brejnholt\altaffilmark{4},
Felix Fuerst\altaffilmark{1},
Carl C. Liebe\altaffilmark{7},
Craig B. Markwardt\altaffilmark{6},
Melania Nynka\altaffilmark{5},
Julia K. Vogel\altaffilmark{2}
Dominic J. Walton\altaffilmark{1},
Daniel R. Wik\altaffilmark{6},
David~M. Alexander\altaffilmark{15}, 
Lynn R. Cominsky\altaffilmark{16},
Ann E. Hornschemeier\altaffilmark{6},
Allan Hornstrup\altaffilmark{4},
Victoria M. Kaspi\altaffilmark{12},
Greg M. Madejski\altaffilmark{11},
Giorgio Matt\altaffilmark{17},
Silvano Molendi\altaffilmark{18},
David M. Smith\altaffilmark{19},
John A. Tomsick\altaffilmark{3},
Marco~Ajello\altaffilmark{3}, 
David R. Ballantyne\altaffilmark{20},
Mislav Balokovi\'{c}\altaffilmark{1}, 
Diddier Barret\altaffilmark{13}, 
Franz~E. Bauer\altaffilmark{21},
Roger D. Blandford\altaffilmark{9},
W. Niel Brandt\altaffilmark{22,23}, 
Laura W. Brenneman\altaffilmark{24},
James Chiang\altaffilmark{9},
Deepto Chakrabarty\altaffilmark{25},
Jerome Chenevez\altaffilmark{4},
Andrea Comastri\altaffilmark{26},
Francois Dufour\altaffilmark{12},
Martin Elvis\altaffilmark{24},
Andrew~C. Fabian\altaffilmark{27},
Duncan Farrah\altaffilmark{28},
Chris L. Fryer\altaffilmark{29},
Eric~V. Gotthelf\altaffilmark{5},
Jonathan E. Grindlay\altaffilmark{24},
David J. Helfand\altaffilmark{30},
Roman Krivonos\altaffilmark{3},
David L. Meier\altaffilmark{7},
Jon M. Miller\altaffilmark{31},
Lorenzo Natalucci\altaffilmark{32},
Patrick Ogle\altaffilmark{33},
Eran O. Ofek\altaffilmark{34},
Andrew Ptak\altaffilmark{6},
Stephen P. Reynolds\altaffilmark{35},
Jane R. Rigby\altaffilmark{6},
Gianpiero Tagliaferri\altaffilmark{36},
Stephen E. Thorsett\altaffilmark{37},
Ezequiel Treister\altaffilmark{38},
C.~Megan Urry\altaffilmark{39}
}


\altaffiltext{1}{Cahill Center for Astronomy and Astrophysics, California Institute of Technology, Pasadena, CA 91125; fiona@srl.caltech.edu.  }
\altaffiltext{2}{Lawrence Livermore National Laboratory, Livermore, CA
94550}
\altaffiltext{3}{Space Sciences Laboratory, University of California, Berkeley, CA 94720}
\altaffiltext{4}{DTU Space - National Space Institute, Technical
University of Denmark, Elektrovej 327, 2800 Lyngby, Denmark}
\altaffiltext{5}{Columbia Astrophysics Laboratory, Columbia University, New York, NY 10027}
\altaffiltext{6}{NASA Goddard Space Flight Center, Greenbelt, MD 20771}
\altaffiltext{7}{Jet Propulsion Laboratory, California Institute of Technology, Pasadena, CA 91109}
\altaffiltext{8}{ASI Science Data Center, c/o ESRIN, via G. Galilei, I-00044 Frascati, Italy}
\altaffiltext{9}{INAF - Osservatorio Astronomico di Roma, via di Frascati 33, I-00040 Monteporzio, Italy}
\altaffiltext{10}{RIKEN, 2-1 Hirosawa, Wako, Saitama, 351-0198, Japan}
\altaffiltext{11}{Kavli Institute for Particle Astrophysics and Cosmology, SLAC National Accelerator Laboratory, Menlo Park, CA 94025}
\altaffiltext{12}{Department of Physics, McGill University, Rutherford Physics Building, 3600 University Street, Montreal, Quebec, H3A 2T8, Canada}
\altaffiltext{13}{UniversitŽ de Toulouse; UPS-OMP; IRAP; Toulouse, France \& CNRS; Institut de Recherche en Astrophysique et Plan\'etologie; 9 Av. colonel Roche, BP 44346, F-31028 Toulouse cedex 4, France}
\altaffiltext{14}{Inter-University Center for Astronomy and
Astrophysics, Post Bag 4, Ganeshkhind, Pune 411007, India}
\altaffiltext{15}{Department of Physics, Durham University, Durham DH1 3LE, UK}
\altaffiltext{16}{Department of Physics and Astronomy, Sonoma State University, Rohnert Park, CA 94928}
\altaffiltext{17}{Dipartimento di Matematica e Fisica, Universit\`a Roma Tre, via della Vasca Navale 84, 00146 Roma, Italy}
\altaffiltext{18}{IASF-Milano, INAF, Via Bassini 15, I-20133 Milano, Italy}
\altaffiltext{19}{Physics Department and Santa Cruz Institute for Particle Physics, University of
      California Santa Cruz, Santa Cruz, CA 95064}
\altaffiltext{20}{Center for Relativistic Astrophysics, School of Physics, Georgia Institute of Technology, Atlanta, GA 30332}
\altaffiltext{21}{Pontificia Universidad Cat\'{o}lica de Chile, Departamento de Astronom\'{\i}a y Astrof\'{\i}sica, Casilla 306, Santiago 22, Chile}
\altaffiltext{22}{Department of Astronomy and Astrophysics, The Pennsylvania State University, 525 Davey Lab, University Park, PA 16802}
\altaffiltext{23}{Institute for Gravitation and the Cosmos, The Pennsylvania State University, University Park, PA 16802}
\altaffiltext{24}{Harvard-Smithsonian Center for Astrophysics, 60 Garden St., Cambridge, MA 02138}      
\altaffiltext{25}{Kavli Institute for Astrophysics and Space Research, Massachusetts Institute of Technology, Cambridge, MA 02139}
\altaffiltext{26}{INAF-Osservatorio Astronomico di Bologna, via Ranzani 1, 40127, Bologna, Italy}
\altaffiltext{27}{Institute of Astronomy, Madingley Road, Cambridge CB3 0HA, UK}
\altaffiltext{28}{Department of Physics, Virginia Tech, Blacksburg, VA, 24061}
\altaffiltext{29}{CCS-2, Los Alamos National Laboratory, Los Alamos, NM 87545}
\altaffiltext{30}{Quest University Canada, Squamish BC V8B 0N8 CANADA }
\altaffiltext{31}{Department of Astronomy, The University of Michigan, 500 Church Street, Ann Arbor, MI, 48109}
\altaffiltext{32}{Istituto di Astrofisica e Planetologia Spaziali, INAF, Via Fosso del Cavaliere 100, Roma, I-00133, Italy} 
\altaffiltext{33}{Infrared Processing and Analysis Center, Caltech, Pasadena, CA 91125}
\altaffiltext{34}{Benoziyo Center for Astrophysics, Weizmann Institute of Science, 76100 Rehovot, Israel.}
\altaffiltext{35}{Physics Dept., NC State U., Raleigh, NC 27695}
\altaffiltext{36}{INAF - Osservatorio Astronomico di Brera, Via Bianchi 46, 23807 Merate, Italy}
\altaffiltext{37}{Department of Physics, Willamette University, Salem, OR 97301}
\altaffiltext{38}{Universidad de Concepci\'{o}n, Departamento de Astronom\'{\i}a, Casilla 160-C, Concepci\'{o}n, Chile}
\altaffiltext{39}{Department of Physics and Yale Center for Astronomy and Astrophysics, Yale University, New Haven, CT 06520-8120}

\begin{abstract}
The {\em Nuclear Spectroscopic Telescope Array (NuSTAR)} mission, launched on 13 June  2012, is the first focusing high-energy X-ray
telescope in orbit.  \nustar\ operates in the band from 3 -- 79~keV, extending the sensitivity of focusing far beyond the $\sim$10~keV
high-energy cutoff achieved by all previous X-ray satellites.     The inherently low-background associated with concentrating
the X-ray light enables \nustar\ to probe the hard X-ray sky with a more than one-hundred-fold improvement in sensitivity over 
the collimated or coded-mask instruments that have operated in this bandpass.    Using its unprecedented combination of sensitivity,
spatial and spectral resolution,  \nustar\ will pursue five primary scientific objectives:
1) probe obscured AGN activity out to the peak epoch of galaxy assembly in the universe (at $z \simless 2$) by surveying selected regions of the sky; 2) 
study the population of hard X-ray emitting compact objects in the Galaxy by mapping the central regions of the Milky Way;
3) study the non-thermal radiation in young supernova remnants, both the hard X-ray continuum and the emission from the radioactive
element $^{44}$Ti; 4) observe blazars contemporaneously with ground-based radio, optical and TeV telescopes, as
well as with {\em Fermi}  and {\em Swift}, to constrain the structure of AGN jets; and, 5) observe line and continuum emission from core-collapse supernovae in the Local Group, and from nearby Type~Ia events, to constrain explosion models.   During its baseline
two-year mission,  \nustar\ will
also undertake a broad program of targeted observations.  The observatory consists of two co-aligned grazing-incidence
X-ray telescopes pointed at celestial targets by a three-axis stabilized spacecraft.   Deployed into a 600~km, near-circular, 6\deg\ inclination orbit,
the Observatory has now completed commissioning, and is performing  consistent
with pre-launch expectations.  \nustar\  is  now executing its primary  science mission, and with an expected orbit lifetime of
ten years, we anticipate proposing a guest investigator program, to begin in Fall 2014.
\end{abstract}


\section{Introduction}

The last decade has seen a major technological advance in the ability to efficiently focus hard X-rays/soft gamma-rays. This breakthrough has enabled the development of instruments that are orders of magnitude more sensitive compared to the collimators and coded-mask cameras previously used to observe the cosmos at the these energies.
Focusing instruments achieve large concentration factors, such that their collecting
area is significantly larger (by factors of 1000 or greater) than the detector area used to register the signal.   In the hard X-ray
band, where particle interactions result in high detector backgrounds, large concentration factors result in enormous improvements in the signal-to-background
ratio over coded mask cameras, where telescope effective areas are typically less than ($\sim$50\%) of the detector area.   
Focusing telescopes operating at energies above 10~keV have been developed and deployed on the High-Energy Focusing Telescope (HEFT),
High-Energy Replicated Optics (HERO) and InFOCUS balloon platforms \cite{hcc+05,raa+02,tko+05}.
These balloon experiments paved the way for \nustar, ART-XC (to be deployed on {\em Spectrum-Roentgen-Gamma}) and
{\em ASTRO-H}.

The {\em Nuclear Spectroscopic Telescope Array (NuSTAR)} Small Explorer mission is the first astronomical telescope in orbit to 
utilize the new generation of hard X-ray optics and solid-state detector technologies to carry out high-sensitivity 
observations at X-ray energies significantly greater than 10~keV.  
The {\em NuSTAR} instrument, which focuses X-rays in the band from 3 -- 79 keV, is based in large part on the technologies developed for HEFT).  
\nustar\ began its detailed design phase in February 2008,  and was
launched into a near-equatorial, low-Earth orbit just over four years later, on 13 June  2012.   

In its two-year primary science mission, {\em NuSTAR} will undertake a broad range of  programs that emphasize its
unique combination of hard X-ray sensitivity,  sub-arcminute angular resolution, and sub-keV spectral response.
In this paper we provide an overview of the mission (\S~2), describe the mission and instrument designs (\S~3,4), summarize
the key in-flight calibrations and performance (\S~5),  describe the ground data systems (\S~6), target of opportunity response (\S~7),
data analysis approach (\S~8), and education and public outreach program (\S~9).  We also describe the motivation for observations
to be undertaken as part of the two-year baseline science mission 

\section{ {\em NuSTAR} Overview}

\begin{figure*}[htbp]
  \centering
  \includegraphics[scale=0.5]{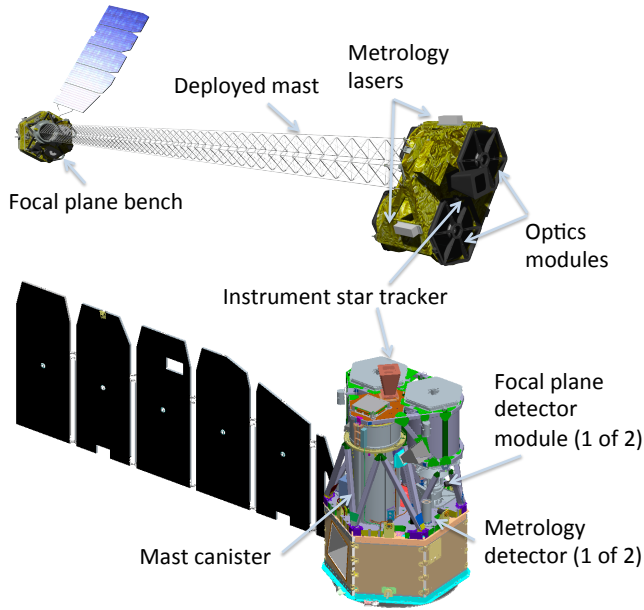}
  \caption{Diagram of the observatory in the stowed (bottom) and deployed (top) configurations.}
  \label{fig-observatory}
\end{figure*}

The \nustar\ observatory, shown in Figure~\ref{fig-observatory}, consists of two co-aligned hard X-ray telescopes which 
are pointed at celestial targets by a three-axis-stabilized spacecraft.   The observatory was launched from the Reagan 
Test Site on the Kwajalein Atoll in the South Pacific in a compact, stowed configuration
on a Pegasus XL vehicle.  On the ninth day after launch an extendible mast was deployed to achieve the 10.14~m instrument focal length.
Table~1 provides an overview of key mission and orbit parameters.   The launch site was chosen in order to
deploy into a low-inclination orbit, where passages through the South Atlantic Anomaly (SAA) -- a region with a high concentration of
trapped particles -- are minimized.   The observatory has no consumables, with the lifetime limited only by the lack of redundant
systems and the orbit decay which will result in re-entry in about 10 years.

\begin{deluxetable}{lc}
\tablewidth{0pt}
\tablecaption{{\it NuSTAR} Mission Parameters.}\label{table.mission}
\tablehead{
\colhead{Mission Parameter} &
\colhead{Value}}
\startdata
Mass 		  		& 350 kg \\
Power 		  		& 600 W \\
Orbit 		  		& Low Earth, $650 \times 610$ km \\
Orbit inclination 		& 6\deg \\
Orbit lifetime 	  		& $\sim$10 yr \\
\enddata
\end{deluxetable}

\begin{figure}[htbp]
  \centering
  \includegraphics[scale=0.35]{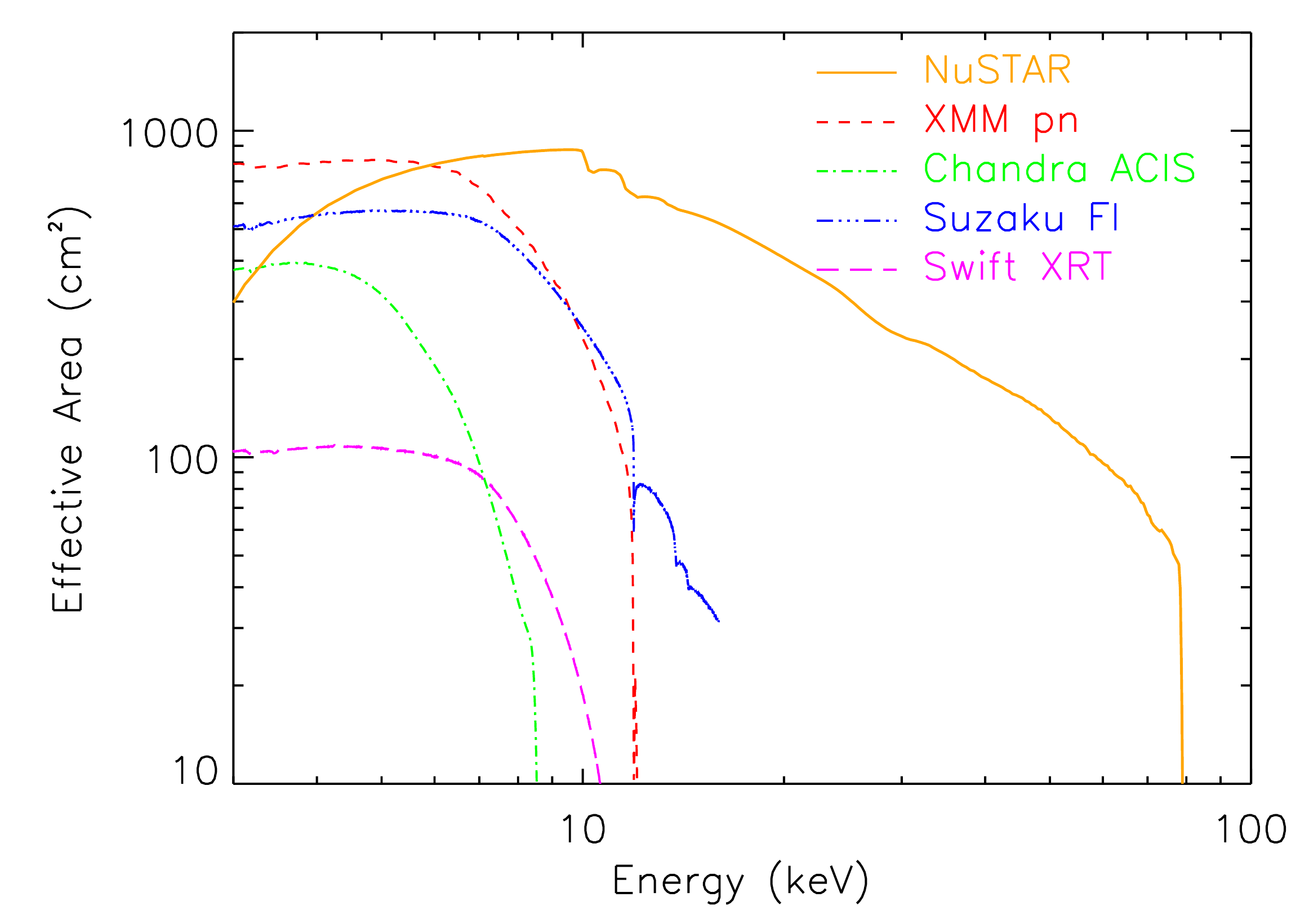}
  \caption{Effective collecting area of \nustar\ compared to selected operating focusing telescopes.  \nustar\ provides good
  overlap with these soft X-ray observatories, and extends focusing capability up to 79~keV. }
  \label{fig-area}
\end{figure}

The \nustar\ science instrument consists of two depth-graded multilayer-coated Wolter I conical approximation (Petre \& Serlemitsos 1985) X-ray optics which
focus onto two independent solid-state focal plane detectors separated from the optics by a $\sim 10$~m focal length.   The optics and detectors are designed to
be as identical as possible, so that the focal plane images can be co-added to gain sensitivity.  Two benches which support the
optics and focal plane systems respectively are separated by a deployable composite mast.  The mast is not sufficiently stiff to maintain the
required relative alignment of the benches on-orbit.  The \nustar\ instrument therefore incorporates an aspect/metrology system
consisting of a  star camera mounted to
the optics bench as well as two laser metrology units which measure the translation, tip, tilt and clocking between the benches. 
This information is used to reconstruct the instantaneous instrument alignment and 
pointing direction.    The resulting corrections are applied on the ground to correctly back-project  individual X-rays onto the sky.

Figure~\ref{fig-area} shows \nustar's collecting area as a function of X-ray energy compared to selected focusing missions currently in operation.
The combination of the low graze angle X-ray optics and the multilayer coatings enables significant collecting area to be achieved
out to 78.4~keV, the location of the Platinum K absorption edge .  The angular resolution of the observatory
is dominated by the optics, and
is 18\arcsec\ FWHM, with a half power diameter of 58\arcsec.  \nustar's focal plane is designed to achieve good energy
resolution in the hard X-ray range, with a FWHM response of 400~eV at 10~keV and 0.9~keV at 60~keV.  An active anti-coincidence
shield reduces background above 10~keV, so that overall the instrument detection threshold represents more than 
two orders of magnitude improvement over
collimated or coded-aperture instruments that have flown in this band.   
Table~2 summarizes the key performance
parameters.

\begin{deluxetable*}{lc}
\tablewidth{0pt}
\tablecaption{Key Observatory Performance Parameters.}
\tablehead{
\colhead{ Parameter} &
\colhead{Value}}
\startdata
Energy range 								& 3 -- 78.4 keV \\ 
Angular resolution (HPD)  						& 58\arcsec \\ 
Angular resolution (FWHM) 						& 18\arcsec \\ 
FoV (50\% resp.) at 10 keV 						& 10\arcmin \\ 
FoV (50\% resp.) at 68 keV 						& 6\arcmin \\ 
Sensitivity (6 -- 10 keV) [10$^6$ s, 3$\sigma$, $\Delta$E/E = 0.5] 	& $2 \times 10^{-15}$ erg cm$^{-2}$ s$^{-1}$ \\ 
Sensitivity (10 -- 30 keV) [10$^6$ s, 3$\sigma$, $\Delta$E/E = 0.5] 	& $1 \times 10^{-14}$ erg cm$^{-2}$ s$^{-1}$ \\ 
Background in HPD (10 -- 30 keV) 					& $1.1 \times 10^{-3}$ cts s$^{-1}$ \\ 
Background in HPD (30 -- 60 keV) 					& $8.4 \times 10^{-4}$ cts s$^{-1}$  \\ 
Spectral resolution (FWHM) 						& 400~eV at 10~keV, 900~eV at 68 keV \\ 
Strong source ($> 10\sigma$) positioning 				& 1.5\arcsec ($1\sigma$) \\ 
Temporal resolution 							& 2 $\mu$s \\ 
Target of opportunity response 						& $< 24$ hr \\ 
Slew rate								& 0.06\deg~s$^{-1}$ \\
Settling time                                                             & 200~s (typ) \\
\enddata
\label{table.perf}
\end{deluxetable*}

\section{\nustar\ Observatory and Mission Design}

 The \nustar\ observatory is pointed at predetermined locations on the sky by a three-axis stabilized spacecraft based on  Orbital Science's LEOStar bus.    \nustar\ is designed for long  observations (1~day - weeks in duration).  The observatory
does not  re-orient during periods of Earth occultation, and  for a typical celestial source ($55$ \deg $  > \delta > -55$\deg) the observing
efficiency is 55\%, including occultations and SAA passages; sources at high latitudes can be observed with close to
90\% efficiency.     

A four-head star camera system, the Technical University of Denmark's $\mu$ASC~\cite{jjd04}, is used to determine both instrument and
spacecraft attitude.    Three of the four units are mounted on or near the spacecraft bus and are combined with other spacecraft  sensors to provide 
attitude
control and determination.  The fourth unit, mounted to the instrument optics bench, is combined with a laser metrology
system to determine instrument pointing and alignment (\S~\ref{sec-attitude}).  As a result of the multi-head tracker design and the lack of thermal
constraints, \nustar\  can perform science observations at any given time over more than 80\% of the sky.  For science targets the 
primary restrictions are a cone of 39\deg\ around the Sun and 14\deg\ around the full Moon, with other small regions excluded by the
requirement that one spacecraft star tracker be available at all times.    \nustar\ can point closer to (or even at) the Sun and Moon;
however, pointing reconstruction is degraded due to the lack of availability of the  optics bench star camera.

\nustar\ was designed such that one side of the observatory always faces the Sun, and pointing to a celestial target is 
achieved by rotating the observatory about the Sun-Earth vector.  This  allows the use of a solar array with a single axis of rotation 
and simplifies the thermal design.  As a consequence, the observatory position angle 
is restricted to $0 - 10$\deg~from the Sun at any given point in the year,  so that the position angle for a
particular target depends on when during the year it is observed.  
Science target observations are generally performed in an inertial pointing mode which keeps the position angle fixed. A small slew is 
performed every few days to keep the solar array optimally pointed at the Sun.

\section{The \nustar\ Science Instrument}

\begin{figure}[htbp]
  \centering
  \includegraphics[scale=0.5]{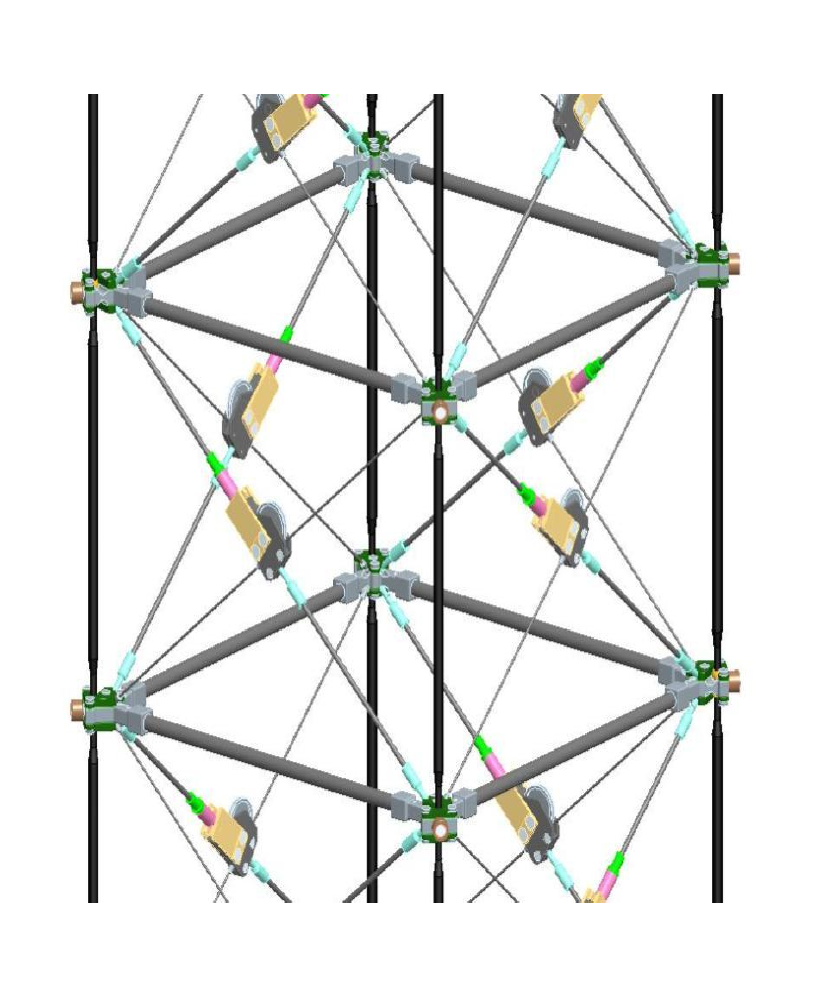}
  \caption{Close up including one full bay of the \nustar\ mast.  Diagonal cables tension and latch during deployment, locking
  the  carbon fiber elements into a stiff structure. }
  \label{fig-mast}
\end{figure}

The two co-aligned hard X-ray grazing incidence telescopes form the core of the \nustar\ instrument.   The two optics modules are 
mounted, along with one of the star tracker heads, to a composite, thermally stable bench (see Figure~\ref{fig-observatory}).    The shielded focal plane modules
are mounted to an aluminum structure which is attached to the spacecraft.   The two benches are separated by a mast which was deployed
after launch.  The mast consists of 57 rectangular structures made stiff after deployment by diagonal cables that latch as the
system is deployed (see Figure~\ref{fig-mast}).  
Due to thermal conditions that vary over an orbit and with aspect angle relative to the Sun, the alignment of the optics
and focal plane benches  changes in translation, tip, tilt and relative rotation during an observation.   These changes  move the location of the 
optical axis and also the X-ray spot on the detector by about 3~mm (1\arcmin ) each.  This changes the alignment 
and vignetting functions as a function of time.  A laser metrology
system, combined with the optics-bench-mounted star tracker, measure the varying translation, tilt and rotation of the optics relative
to the detectors.   These measurements are combined during data processing on the ground to remove image blurring and determine correct response files.   
The metrology system is read out at a rate of 16~Hz, and the instrument star tracker head samples at 4~Hz.   Liebe~{\em et al.} (2012)\nocite{lck+12}
describe the metrology system in detail, and Harp~{\em et al.} (2010)\nocite{hlc+10}
provide details of the instrument alignment and pointing reconstruction.

\subsection{Optics}

\begin{deluxetable*}{lccc}
\tablewidth{0pt}
\tablecaption{Optics Parameters.}
\tablehead{
\colhead{Optics Parameter} &
\colhead{Value} &
\colhead{Optics Parameter} &
\colhead{Value}} 
\startdata
Focal length    	& 10.14 m 		& Shell length     & 22.5 cm \\
\# Shells 		& 133 			& Min. graze angle & 1.34 mrad \\
\# Azimuthal segments  	& 6 (inner)/12 (outer)  & Max. graze angle & 4.7 mrad \\
Inner radius 		& 5.44 cm 		& Coatings (inner) & Pt/C \\
Outer radius 		& 19.1 cm 		& Coatings (outer) & W/Si  \\ 
\enddata
\label{tab-optics}
\end{deluxetable*}

The two \nustar\ optics  modules each contain 133 nested multilayer-coated grazing incidence shells in a conical approximation to
a Wolter-I geometry.   Table~3 provides the key parameters. Each shell is comprised of either 12 or 24 formed
glass segments, depending on the radius in the optic.  The glass is thin (0.2~mm) sheet glass manufactured by Schott. The optics are coated with depth-graded multilayer structures which, for energies above $\sim 15$~keV, increase 
the graze angle for which significant reflectance can be achieved.   This increases the field of view (FoV) and high-energy 
collecting area relative to standard metal coatings.    The coating materials and prescriptions vary as a function graze angle (or, equivalently, shell radius) and have been designed to optimize the broad-band energy response and FOV.  The inner 89 shells are coated with depth-graded Pt/C multilayers that reflect efficiently below the Pt K-absorption edge at 78.4 keV.  The outer 44 shells are coated with depth-grade W/Si multilayers that reflect efficiently below the W 
K-absorption edge at 69.5 keV.

\begin{figure}[htbp]
  \centering
  \includegraphics[scale=0.18]{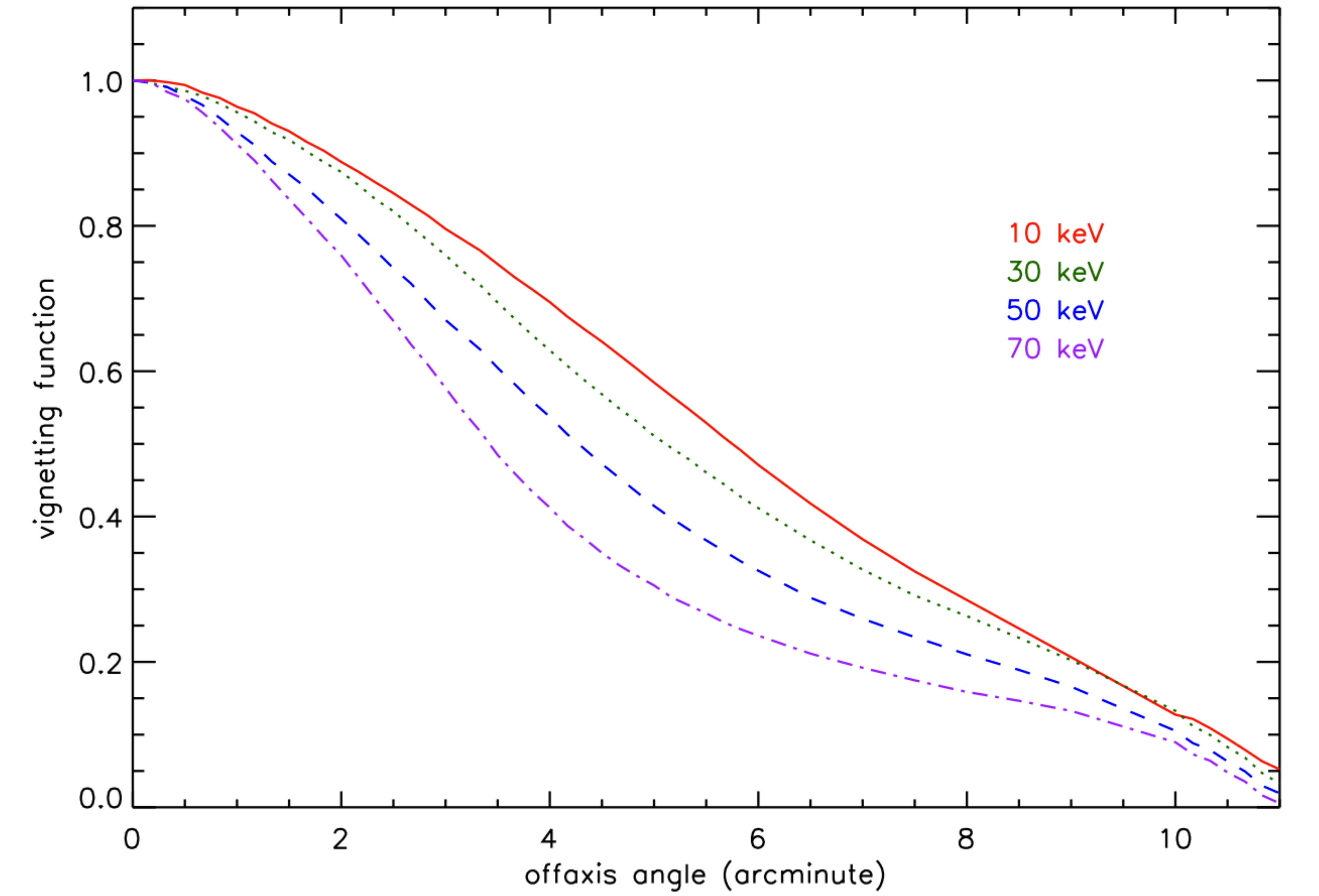}
  \caption{The reduction in effective area as a function of off-axis angle resulting from vignetting in the optics.   The curves
  show the vignetting function for different incident energies.  Due to the tightly nested low-graze-angle design and the
  change in reflectance of the shells as a function of energy, the vignetting is strongly energy dependent above 15 keV.}
  \label{fig-vignetting}
\end{figure}

Compared to soft X-ray telescopes, the optics prescription utilizes smaller graze angles (the angle between an on-axis incident X-ray and the optics shell).  This means vignetting losses will become noticeable for off-axis positions below 10\arcmin.  This factor, when combined the dependence of the multilayer reflectivity on both the incident angle and energy of the X-ray (higher energy photons have higher reflectivity at shallower graze angles), results in a complex behavior for the effective area of the optic.  As illustrated in Figure~\ref{fig-vignetting}, the optics area as a function of off-axis source position decreases with increasing photon energy.  The effective FOV, defined by the furthest off-axis position that has 50\% of the on-axis effective area,  will therefore
decrease with energy:  it is 10\arcmin\ at 10 keV, and 6\arcmin\ at 68 keV.

The  optics angular response is dominated by figure errors inherent in the substrates and in the mounting technique.   As a result, unlike {\em Chandra}, 
where the response is dictated by the optical design, for \nustar\ 
the point spread function is not a strong function of off-axis angle.     While the detailed shape changes, to first order the area of the encircled energy 
contours remain approximately constant with off-axis angle.

A detailed description of the optics can be found in 
Hailey~{\em et al.} (2010)\nocite{hab+10},
the coatings are described in Christensen~{\em et al.} (2011)\nocite{cjb+11}, and the fabrication approach is detailed in Craig~{\em et al.} (2011)\nocite{cab+10}. 

\subsection{Focal Plane} 

\begin{figure}[htbp]
  \centering
  \includegraphics[scale=0.6]{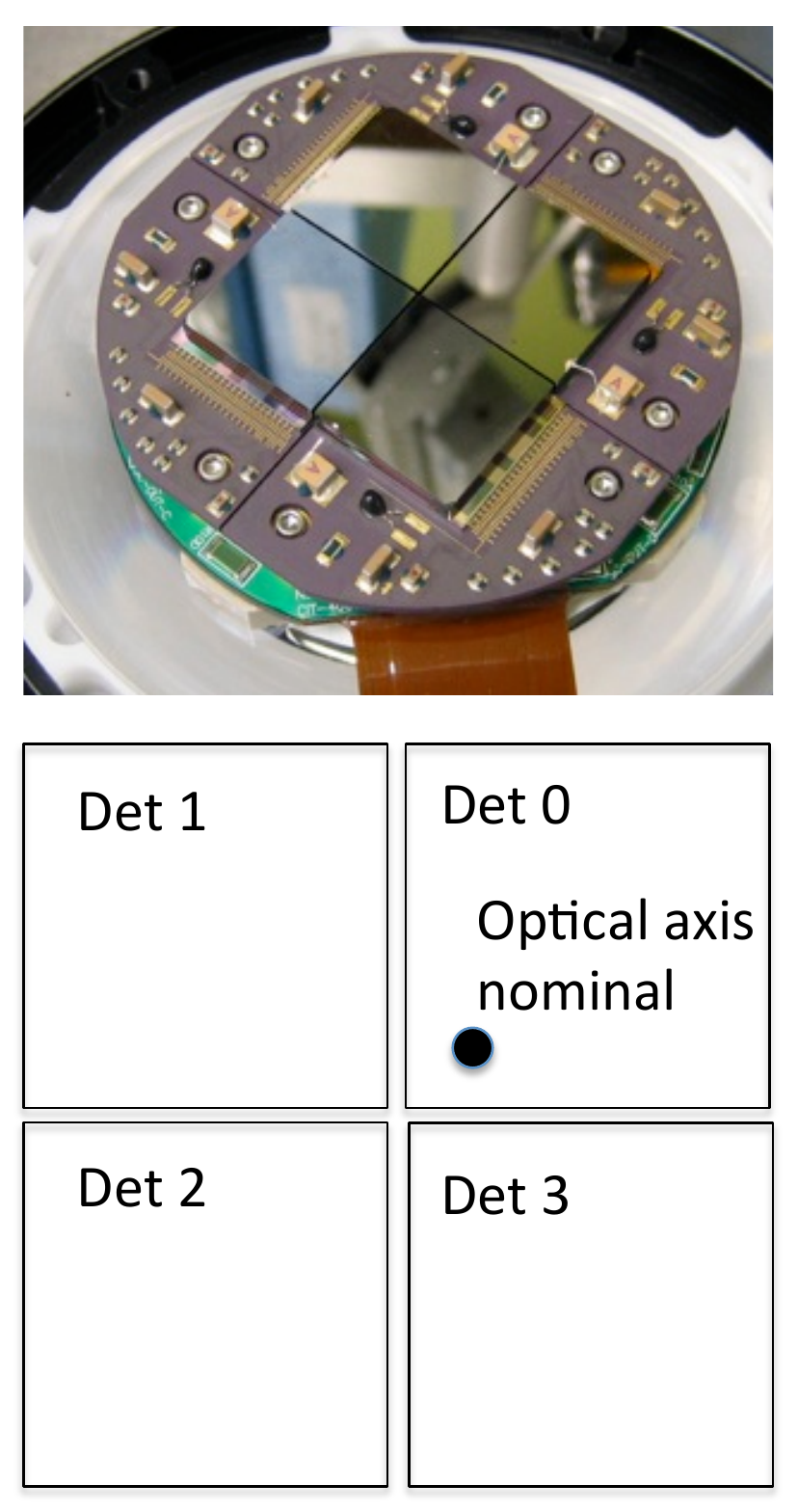}
  \caption{Photo of a \nustar\ focal plane (top).   The focal plane consists of a 2 $\times$ 2 array of
  CdZnTe pixel detectors, each with 32 $\times$ 32 pixels 0.6~mm in size.   The detectors are referred
  to as Det 0 -- 3.   The optical axis nominal location is approximately 1\arcmin\ from the chip gaps on
  Det 0 (bottom). }
  \label{fig-focalplane}
\end{figure}

Each telescope has its own focal plane module, consisting of a solid state CdZnTe pixel detector (Harrison {\em et al.} 2010) 
surrounded by a CsI anti-coincidence
shield.   Table~4 provides the key focal plane parameters.  A two-by-two array of detectors (Figure~\ref{fig-focalplane}), each with an array of $32 \times 32$,
0.6~mm pixels (each pixel subtending 12.3\arcsec ) provides a 12\arcmin\ field of view.   
In \nustar\ each pixel has an independent discriminator, and individual X-ray interactions trigger the readout process.    On-board
processors, one for each telescope, identify the row and column with the largest pulse height, and read out pulse height information from this pixel
as well as its eight neighbors.   The event time is recorded to an accuracy of 2~$\mu$s relative to the on-board clock.
The event location, energy, and depth of interaction in the detector are computed
from the nine-pixel signals (see Rana~{\em et al.} 2009 and Kitaguchi et al. 2011 for details)\nocite{rch+09,kgh+11}.   
The depth-of-interaction measurements allow an energy-dependent cut to be made to reduce internal detector background.   

Since the  detectors do not employ an integrating CCD-style readout, pulse pileup will not occur until source fluxes of $\sim$10$^{5}$ cts/s/pixel.  
The processing time per event is 2.5~ms, limiting the rate at which events can be read out to between 300 and 400 evts/s/module.    
For each event,  the live time since the previous event is recorded to an accuracy of 1~$\mu$s, so that fluxes can be measured to 1\% 
accuracy even for incident count rates of 10$^4$~cps.  Harrison {\em et al.}  (2010) contains a detailed description of the operation of the custom
ASIC readout.

The focal plane detectors are placed inside a CsI anti-coincidence shield. Events resulting in a simultaneous energy deposition in both the surrounding 
anti-coincidence shield and the detector are rejected on-board by a processor as background.   The opening angle of the shield is large (15\deg\ FWZI),
greatly exceeding the sky solid angle blocked by the optics bench; thus, to limit diffuse background impinging on the detector, a series of aperture stops were deployed
simultaneously with the mast, post-launch.    The aperture stops collimate the detector to a 4\deg\ FWZI field, which is partially but not completely
blocked by the optics bench.   This results in some stray light which dominates the background below 10~keV (\S~\ref{sec-background}).

\begin{deluxetable*}{lccc}
\tablewidth{0pt}
\tablecaption{Focal Plane Parameters.}
\tablehead{
\colhead{Focal Plane Parameter} &
\colhead{Value} &
\colhead{Focal Plane Parameter} &
\colhead{Value}} 
\startdata
Pixel size 	 & 0.6~mm/12.3\arcsec 		& Max. processing rate 			& 400 evt s$^{-1}$~module$^{-1}$ \\
Focal plane size & 12\arcmin $\times$ 12\arcmin & Max. flux meas. rate 			& 10$^4$ cts s$^{-1}$ \\
Hybrid format  	 & 32 pix $\times$ 32 pix   	& Time resolution (relative) 			& 2 $\mu$s \\
Energy threshold & 2 keV 			& Dead time fraction (@ threshold)     & 5\% \\ 
\enddata
\label{tab-focalplane}
\end{deluxetable*}

\subsection{The Mast and Metrology System}
\label{sec-metrology}

\begin{figure}[htbp]
  \centering
  \includegraphics[scale=0.5]{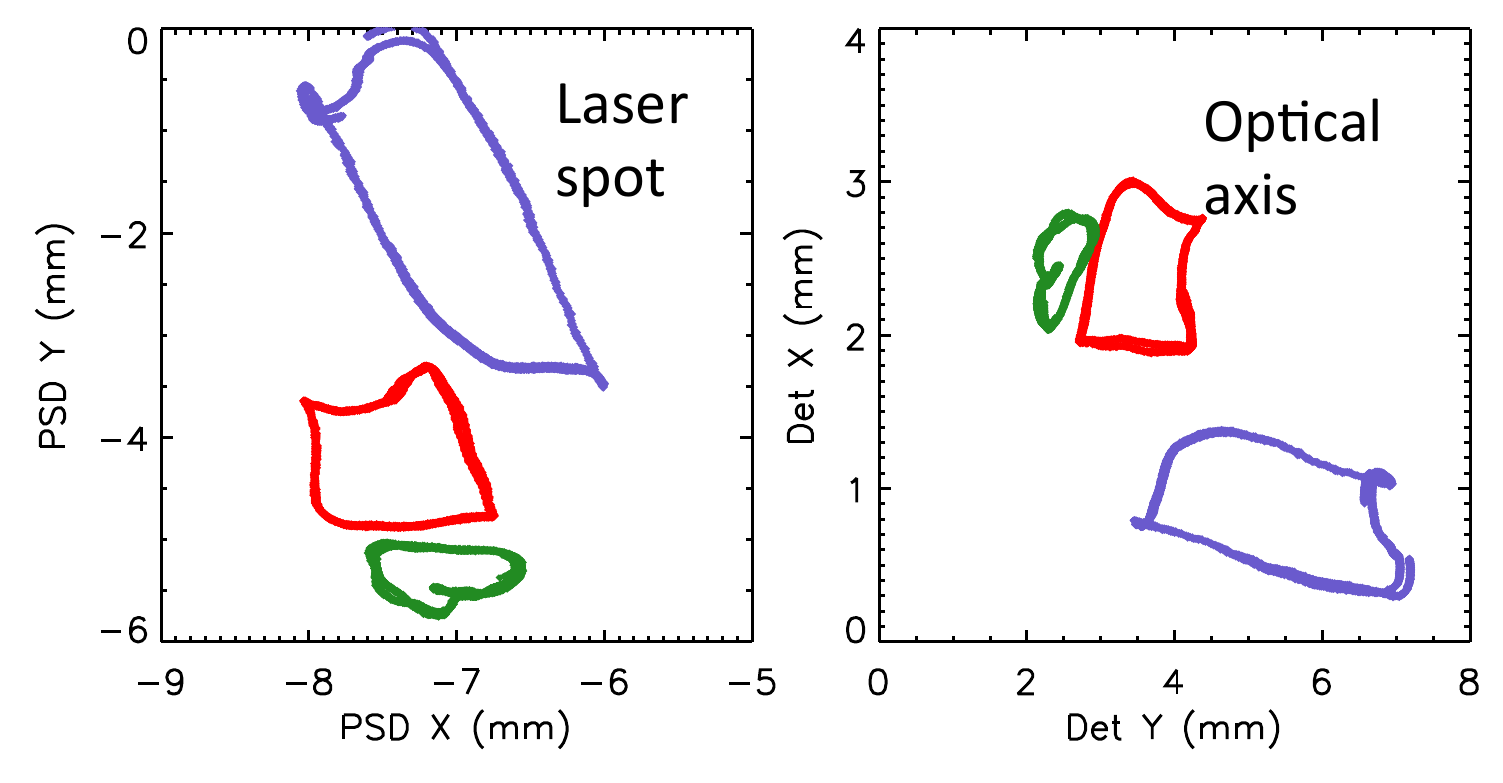}
  \caption{Left: Measured location of the laser beam on the metrology system position sensitive detector (PSD) for module A.  The 
  three sets of tracks show 5-6 orbits each for three different celestial targets differing in pointing direction relative to the Sun.
  The character of the motion changes with thermal conditions,  with the maximum amplitude being $\sim$3~mm.   Right: Motion
  of the Optical Axis (OA) of the X-ray optic on the focal plane detector resulting from the mast motion for module A.  The optical
  axis moves by about 3~mm, or 1\arcmin\  over the course of an orbit}
  \label{fig-metlib}
\end{figure}

The \nustar\ mast structure provides the required 10.14~m separation between
the optics and the focal plane detectors.  The mast was engineered by ATK Space
Systems from segments made up of carbon fiber, aluminum and steel
components, designed to be as isothermal as possible in order to minimize
structural distortions.  The 57 bays (see Figure~\ref{fig-mast}) of the mast were deployed on orbit
and locked in place by tensioned steel cables.   Although the mast
has a low net coefficient of thermal expansion, orbital day/night extremes
create residual deflections that result in motion of the optical axis and the X-ray focal point on
the focal plane detectors.   Thermal modeling of the system pre-lauch
indicated that the  deflection would range from
less than one mm per orbit in the most favorable orientations to a few mm in
the least favorable.  

The mast motion is tracked by the combination of a metrology system~\cite{lck+12} and
the instrument star camera head.  The metrology system consists of two
IR lasers mounted on the optics bench with their beams focused on two
corresponding detectors on the focal plane bench.  The lasers are mounted in a temperature
controlled Invar structure rigidly mounted to the optics bench.   The star
tracker is also mounted to that bench, which was engineered from carbon
fiber structural elements to provide less than 2\arcsec\ of relative deflection
with respect to the optics modules during on-orbit thermal conditions.

The laser beam positions are monitored by Silicon position-sensitive detectors mounted near
the focal plane modules.   Filters and baffles mounted at the detectors
reject scattered Sun and Earth light and provide for high signal-to-noise,
producing $\sim$10~$\mu$m  (0.1\arcsec) accuracy for each laser spot position.  On-ground
analysis of signals from the two detectors, combined with the star tracker
quaternion, separates translational and rotational deflections in the
positions of the two benches and enables both accurate correction of the
position of the X-ray photons on the focal plane modules as well as construction of the
vignetting function appropriate for a given observation.

\section{In-flight Calibration and Science Performance}

During the the first two months on-orbit, the \nustar\ observatory made a series of observations for performance verification,
as well as for internal calibration and cross-calibration with other missions.   The pre-launch  \nustar\ responses were based on
ground calibration data, and they are currently being
adjusted to reflect on-orbit measurements.  In the case of the response files, required post-launch adjustments are at the 10 -- 20\% level
for both the absolute normalization and relative responses. 
These discrepancies are consistent with the overall accuracy of ground calibration, and arise from a combination of statistical and systematic uncertainties in ground measurements and simplified models of complex phenomena associated with high-energy multilayers and the low-energy (E$< $10~keV) detector response.   The cost constraints of a Small Explorer precluded extensive ground calibrations, 
and the final response functions were always planned to be adjusted through in-flight measurements and cross-calibration with other missions.
The current responses are acceptable for some ranges of off-axis angles, and for energies below 10~keV, enabling science
data reduction to proceed.   Current expectations are that calibration will be completed by April 2013.
A more detailed calibration paper is in preparation (Madsen {\em et al.} 2013).  
The sections below briefly present results from cross-calibration and performance verification.

\subsection{Point Spread Function and Optics Response}

\begin{figure}[htbp]
  \centering
  \includegraphics[scale=0.6]{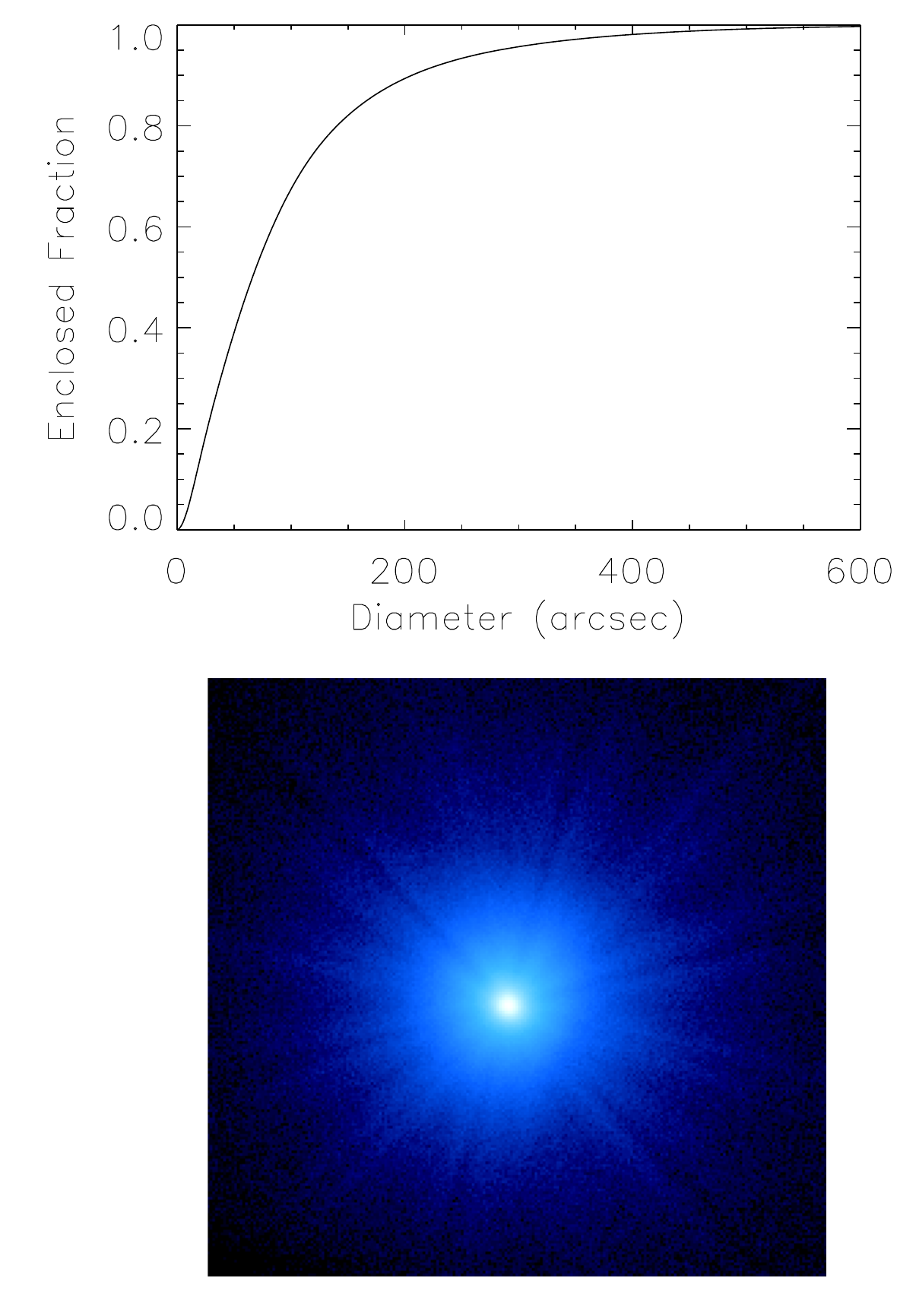}
  \caption{Measurement of the on-axis PSF of module A using data from observations of the bright point source GRS~1915+105.  The
  top panel shows the encircled energy as a function diameter of a circular extraction region.   The bottom panel shows an image stretched
  so that features from the mounting structure are evident, demonstrating the azimuthal symmetry of the response.}
  \label{fig-psf}
\end{figure}

The PSF of the telescopes is a convolution of the
optics and focal plane detector response combined with residual errors from the metrology system and attitude reconstruction.  The dominant contributor is
the optics, with residual terms arising from finite detector sampling and imperfect mast motion and attitude reconstruction at the few-arcsecond
level.  The PSF has been measured in-flight using high signal-to-noise observations of bright point sources.  
The PSF shape consists of a sharp inner core with a full width half maximum of  18\arcsec\ and a half power diameter (diameter of a circle 
enclosing half of the X-ray counts from a point source) of
58\arcsec.  Figure~\ref{fig-psf} shows the encircled counts fraction as a function of diameter for a circular extraction region (left panel) as
measured for telescope module A (the telescope and focal plane modules are designated by  A and B),  from an observation of the accreting Galactic binary
black hole GRS~1915+105.   The right panel of the figure shows a stretched image, demonstrating the azimuthal symmetry of the response.   The PSF for module B is within 2\arcsec\ of that of module A.

\begin{figure}[htbp]
  \centering
  \includegraphics[scale=0.6]{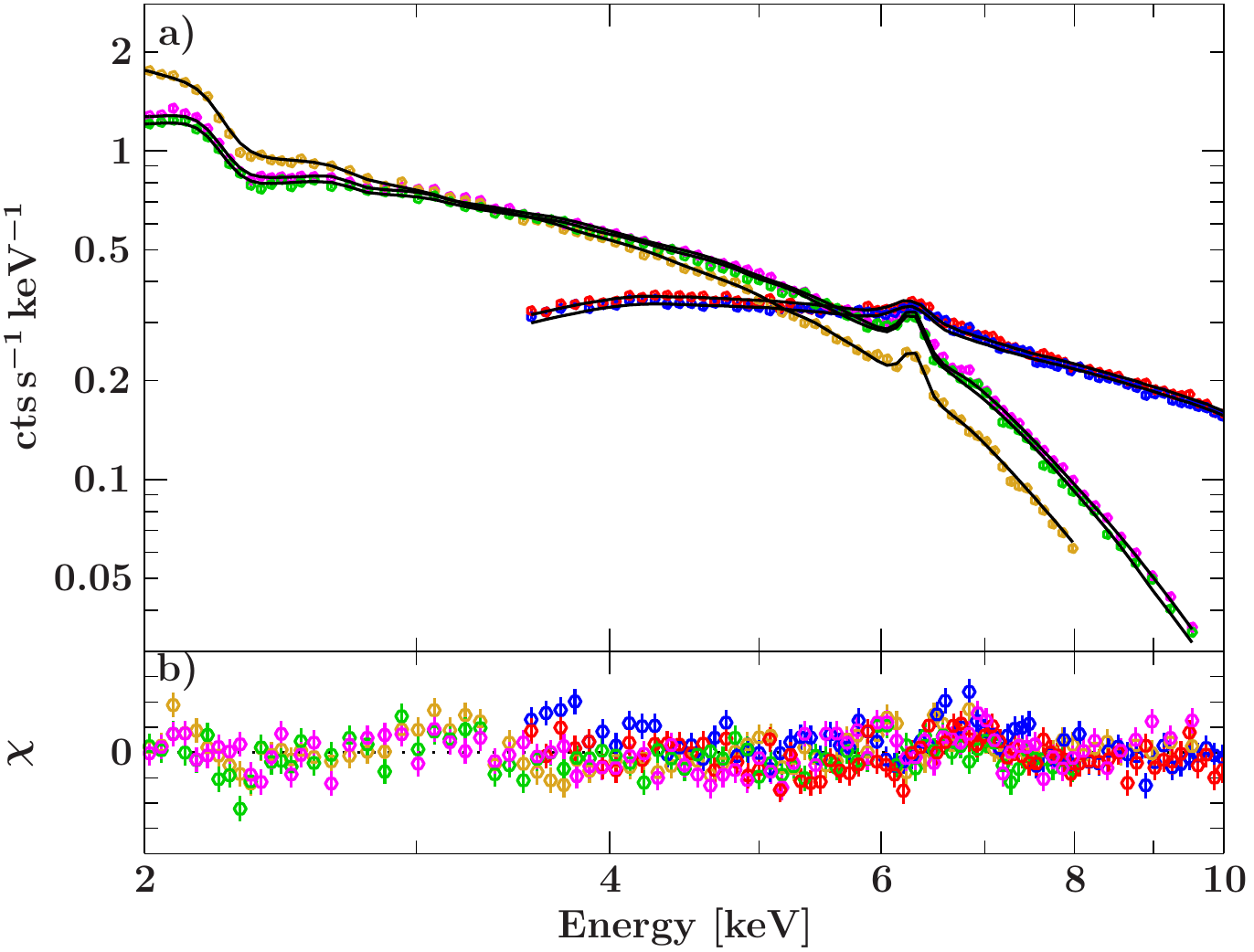}
  \caption{Suzaku XIS and \nustar\ spectra of the AGN IC~4329A in the 2 -- 10 keV band.   The green and orange and magenta
  points are for XIS 0,1 and 3, respectively.   The red and blue points are for \nustar\ focal plane modules A and B.  The top panel shows
  the spectra, with residuals shown as a ratio on the bottom panel.
  The relative calibration of the two instruments is good over this energy range.   The absolute
  cross normalization agrees to within 10\%.}
  \label{fig-IC4329a}
\end{figure}

The normalization of the  \nustar\ effective area in the 2 -- 10~keV band has been
verified in-flight by simultaneous observations of the Crab Nebula and its pulsar, and the
quasar 3C~273 with \xmm, \suzaku, \swift, and \chandra.   Flux measurements in the 5 -- 10~keV band derived for \nustar\
have been compared 
for the indicated calibration observations, as well as for science targets.   We find that the overall
normalization in the 5 -- 10 keV band agrees well with the \xmm~PN, the \suzaku~XIS, and \swift\, with typical cross-calibration corrections
at the  10\% level, similar to the cross-calibration factors among the various soft X-ray observatories.      
The \nustar\ relative response also matches nicely with \xmm~PN and MOS, and \suzaku~XIS in the 3 -- 10 keV band  (see Figure~\ref{fig-IC4329a}).
In a joint observation of the active galaxy IC~4329A with \nustar\ and {\em Suzaku} the absolute flux in the 3 -- 10~keV
band measured with XIS~0 was $8.30\pm0.03 \times 10^{-11}~$erg~cm$^{-2}$~s$^{-1}$, and for  focal plane module A and B 
the measured flux was $9.23\pm0.03 \times 10^{-11}~$erg~cm$^{-2}$~s$^{-1}$  and $8.95\pm0.03 \times 10^{-11}~$erg~cm$^{-2}$~s$^{-1}$, where
quoted errors represent 90\% confidence intervals.   The systematic calibration errors between \nustar's A and B  telescopes are at the
3\% level, with a 10\% normalization uncertainty relative to \suzaku.
Direct comparisons with {\em INTEGRAL}-ISGRI and \suzaku-PIN at higher energies are more difficult, with larger cross-normalization factors
and higher uncertainty in the case of the PIN due to challenges with background subtraction.  However, these cross calibrations are
ongoing, and will be aided by high count-rate joint observations of Cygnus~X-1 and the Crab.

\subsection{Spectral and Temporal Resolution}

\begin{figure}[htbp]
  \centering
  \includegraphics[scale=0.38]{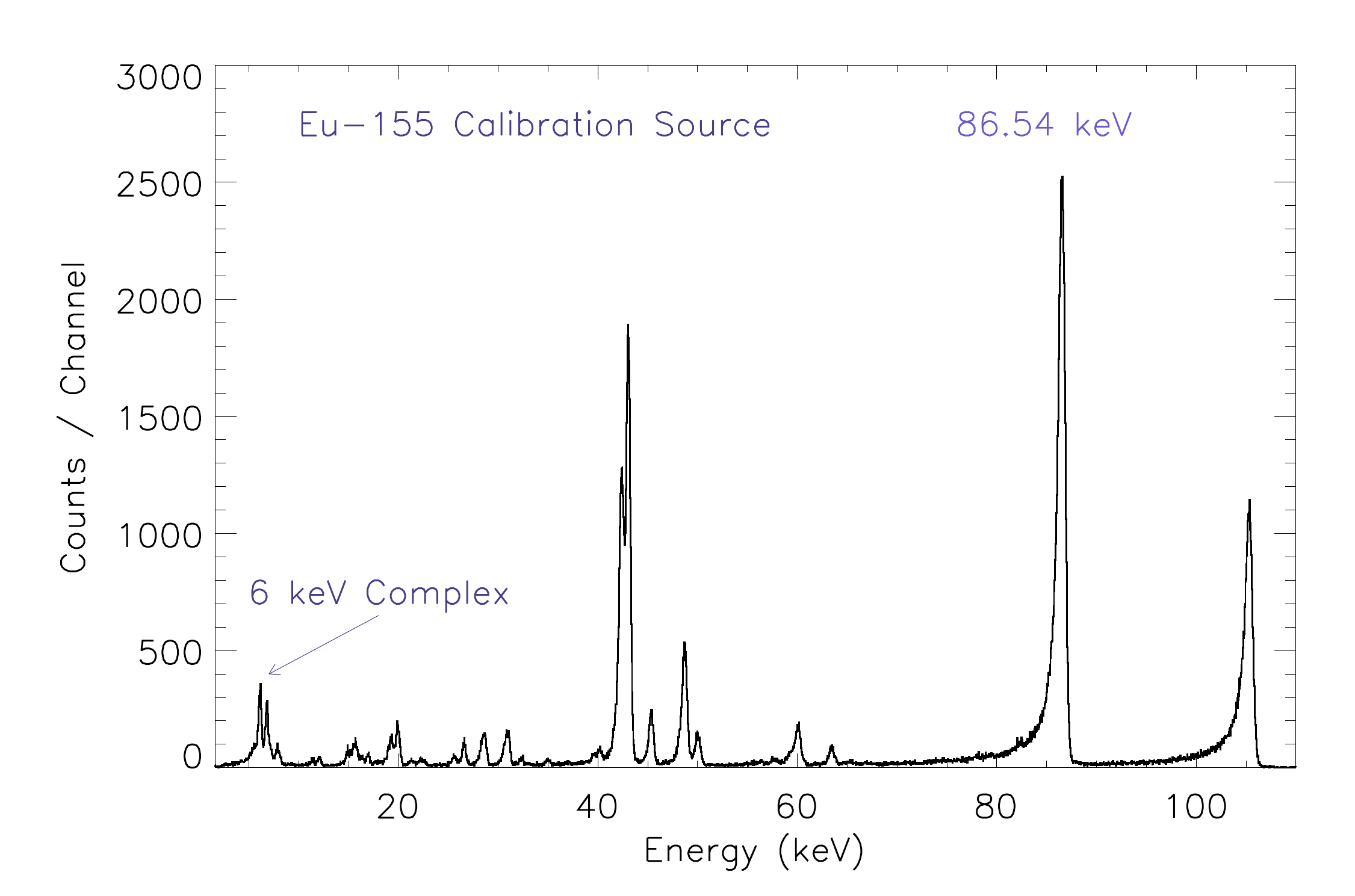}
  \caption{Spectrum of the $^155$Eu flight calibration source for focal plane module A taken during on-orbit commissioning. At energies below
  50 keV the resolution is 400~eV FWHM and is constant with energy, being determined by the electronic noise.   Above this the energy resolution
  broadens due to charge trapping effects in the detector, reaching 1.0~keV FWHM at 86~keV.}
  \label{fig-calsrc}
\end{figure}

\begin{figure}[htbp]
  \centering
  \includegraphics[scale=0.6]{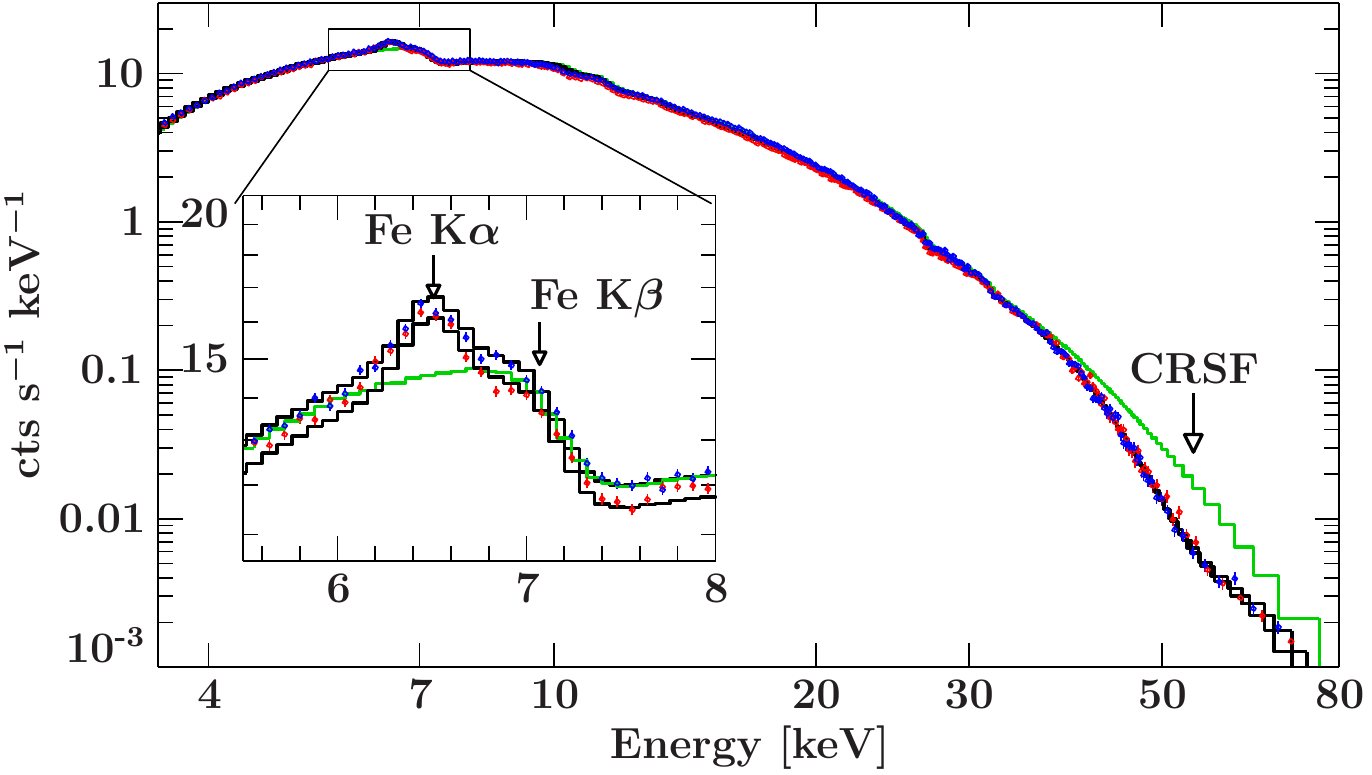}
  \caption{Spectrum with 15~ks exposure of Vela X-1 taken by \nustar .   The red curve shows focal plane module A and the blue curve
  is from B.    The inset shows the Fe-line region.   The good broad band response enables \nustar\ to measure both Fe-line fluxes
  and high energy continuum.   A Cyclotron Resonance Scattering Feature (CRSF) is seen with high significance at 55~keV. }
  \label{fig-vela}
\end{figure}

The in-flight spectral response of the \nustar\ focal plane detectors matches well with pre-launch calibration measurements.
Each focal plane module is equipped with a radioactive $^{155}$Eu calibration source that can be placed in the field of view.   
Figure~\ref{fig-calsrc} shows a spectrum taken during on-orbit commissioning.    At energies below
 $\sim$50 keV the resolution is 400~eV FWHM and is constant with energy, being determined by the electronic noise.   Above this, the energy resolution
 broadens due to charge trapping effects in the detector.  The FWHM response is 1.0~keV at the 86~keV decay line.   The good resolution over a broad
 energy range enables \nustar\ to perform spectroscopy extending from below the Fe-line range up to 
 79~keV.   Figure~\ref{fig-vela} shows a spectrum from a 15~ks exposure of the accreting neutron star Vela X-1 taken by
 \nustar\ during the instrument commissioning phase.     Fe-lines are well measured, along with the continuum extending up 
 to 79~keV.   A clear Cyclotron Resonance Scattering Feature (CRSF) can be seen at 55~keV.

 Because \nustar\  has a triggered readout (similar to a proportional counter and unlike an integrating CCD), timing analysis is
 possible on short timescales for sources of moderate count rate ($< 200$~cts~module$^{-1}$).    For very bright targets the 2.5~ms deadtime
 per event limits the ability to search for features on millisecond timescales.    \nustar\  maintains relative stability of  event timing
 to 1 -- 2 ms by correcting the drift of the spacecraft clock on long (day to weeks) timescales.  This  drift correction is done using
 the relative spacecraft clock time to UT that is routinely measured during ground station contacts.   The absolute
 time relative to UT is currently only known to 5~ms; however, calibration to the sub-millisecond level using a pulsar with a known ephemeris is planned.
 
\subsection{Background}

\label{sec-background}

\begin{figure}[htbp]
  \centering
  \includegraphics[scale=0.45]{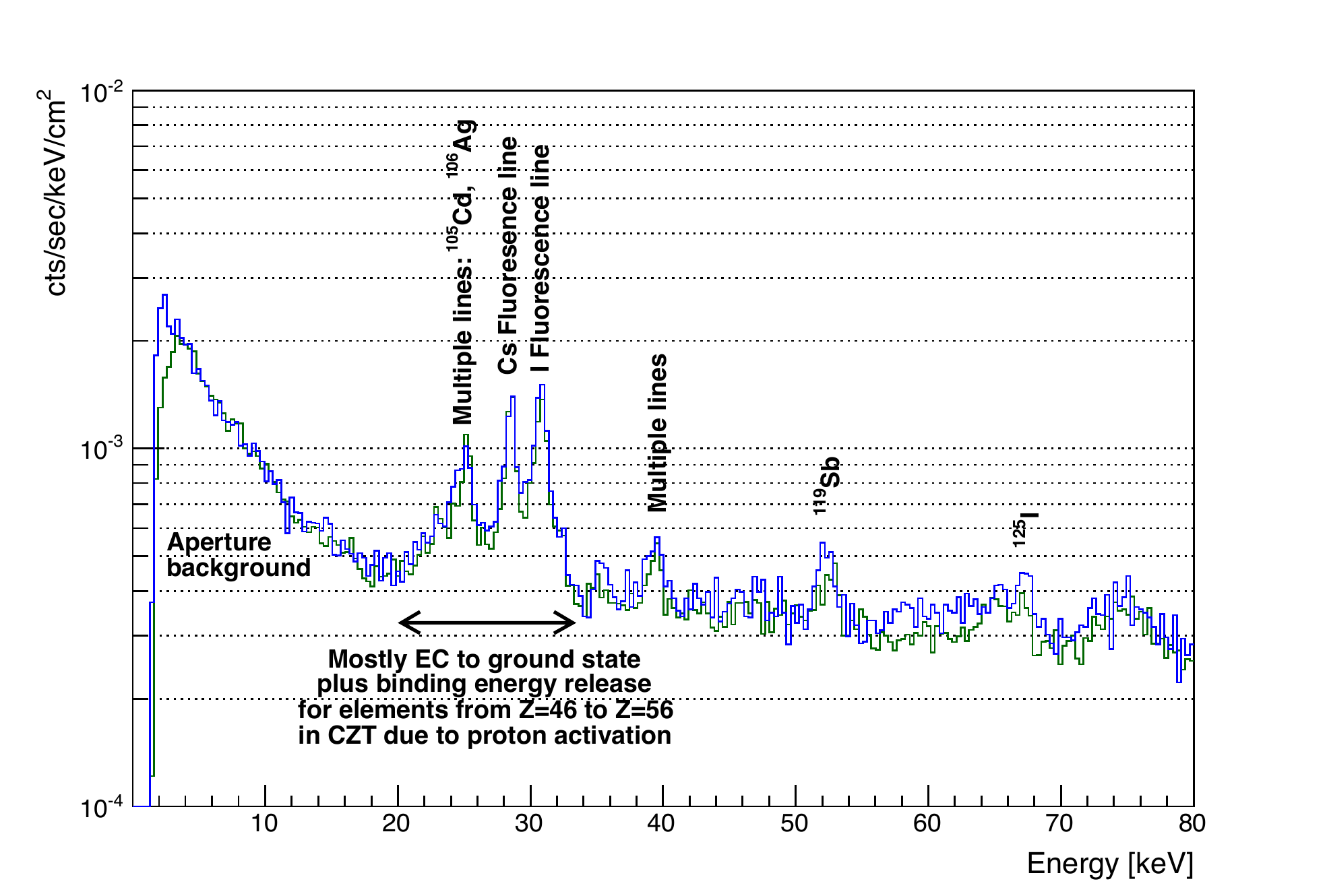}
  \caption{Background spectrum from observations of the Extended Chandra Deep Field South (ECDFS).  At low energies the background
  is dominated by leakage through the aperture stop, while at high energies (E $\simgreat$ 10 keV) atmospheric albedo and activation
  components dominate.  Several prominent lines can be seen in the 20-40 keV band, which are a combination of fluorescence from
  the CsI shield and electron capture to the ground state plus binding energy release from elements with Z= 46 -- 56.}
  
  \label{fig-bkg}
\end{figure}

The  \nustar\ background measured on-orbit is well within the range predicted by pre-launch Monte Carlo modeling.   Background prediction for
a well shielded instrument is difficult due to the poorly characterized input particle spectra, so that pre-launch predictions had uncertainties
of a factor of four.   Figure~\ref{fig-bkg} shows the background spectrum  taken during an observation of the Extended Chandra
Deep Field South in late September 2012.    At low energies (E$<$10~keV)
the background is dominated by diffuse cosmic flux entering through the aperture stop.   This component is not uniform
across the field of view, so that some care is required in choosing regions for background subtraction at low energies
for faint sources.   The internal background
consists of albedo from the Earth's atmosphere, fluorescence lines from the CsI shield, as well as internal activation lines
produced largely by particles trapped in the South Atlantic Anomaly.   This component is spatially uniform across a given detector chip, but 
does vary from detector to detector.

The  low-inclination of the \nustar\ orbit results in stable background compared to missions that experience a larger range of
geomagnetic latitude and more frequent SAA passages.   The background increased slowly in the several months after launch
as a result of activation components, although it has now largely stabilized.   On orbit, the background in the 3 - 79~keV band varies primarily as a function of geomagnetic latitude,  with the exception of the interval from 24 -- 26 keV where two lines
with half-lives of less than one hour result in variations dependent on time elapsed since an SAA passage.  For observations
significantly longer than an orbit, these variations average out, leaving systematic uncertainties in background subtraction of $<1$\%.  

Two modes of background subtraction have been developed depending on whether or not a source-free background region is
available in the field of view.   For point sources with background regions available on the same chip, standard procedures 
employed in soft X-ray focusing telescopes apply.    For extended sources where clear background extraction regions do not
exist, the background can be scaled from deep survey fields and filtered to match the average geomagnetic cutoff for the observation.
Both techniques work well and can be applied to extract spectra at levels well below that of the background.

\subsection{Alignment and Absolute Astrometry}

\label{sec-attitude}

The \nustar\ mast, optics bench and focal plane bench were co-aligned on the ground to within a few millimeters.  This ensured that on orbit
the deployed benches would be well within the adjustment capability of the mast adjustment mechanism mounted at the tip of the mast.  
After deployment on orbit, the mast tip was adjusted in two steps to place the optical axis of the telescopes at the desired position on the focal plane detectors. 
X-ray sources with known positions were then observed to measure the relative positions of the optics, 
focal plane detectors, laser metrology system, and optics-bench-mounted star tracker.  

As described in \S\ref{sec-metrology}, 
the metrology system  measures 
the relative positions of the optics and focal plane benches as a function of orbit, pointing direction, and season, while the star 
tracker provides absolute celestial reference.  Due to thermal effects, the benches move relative to one another in translation over an 
orbit which, when combined
with the mast tilt, results in motion of the optical axis on the X-ray focal plane detectors by 3 -- 4~mm (1 -- 1.2\arcmin -- see
Figure~\ref{fig-metlib}).  This relative motion is measured and removed using the star tracker and metrology system.  For
absolute alignment, we used a set  of reference observations to set the alignment.  The absolute astrometric positions 
of bright X-ray sources with known positions are observed to be within $\pm 5$\arcsec (90\%).  Further updates to the 
alignment parameters are expected to result in a final performance (systematic contribution) of $\pm 3$\arcsec,
with the error being dominated by the inherent stability and accuracy of the optics-bench-mounted star tracker. 

\section{Ground Data Systems}

The primary \nustar\ ground station is located in Malindi, Kenya, and is operated by the Italian Space Agency (ASI).   This station
provides 12-min duration contact opportunities every orbit.   Malindi provides \nustar\ with four
of these contacts per day.  For sources with nominal count rates, all data can easily be downlinked.  For high-count-rate targets ($\simgreat$50 cps),
additional downlink is required from a station located in Hawaii, operated by the Universal Space Network (USN).   Owing to the relatively 
high latitude of the Hawaii station, only three long-duration (6 -- 8 minute) passes per day are possible.   
Commands can be sent to the spacecraft via Malindi, USN Hawaii, or via the Tracking and Data Relay Satellite System (TDRSS). 
This enables almost-continuous command capability, so that the time for responding to targets of opportunity can be
quite short (a few hours). 

From Malindi, data are sent via internet to NASA JSC and then relayed to the Mission Operations Center (MOC) at UCB.
After being unpacked and entered in a database, the Level 0 data are sent by internet to the Science Operations Center (SOC)  located at Caltech, where
they are processed into higher-level products .   This process is typically completed within an hour of a ground station pass,
so that data can be quickly assessed for quality and provided to the science team.

Science planning takes place at the SOC, where observation sequences are prepared and sent via a messaging system to
the MOC.  The MOC converts the observation sequence into an Absolute Time Sequence (ATS) file that is uploaded
to the spacecraft for execution.

\section{Target of Opportunity Response}

\nustar's science program includes a number of Target of Opportunity (ToO) observations which require repointing the
observatory at new or variable phenomena on a timescale of hours to days.  The response time to a ToO is guaranteed to 
be $<24$~hours, and can be considerably less for high priority targets.    After receipt of a ToO request, the SOC must
create a message file in order to reorient the observatory.  This typically can be completed in 1 -- 2 hours, and validation of the request
and preparation for uplink accomplished in an additional 1 -- 2 hours.  If urgent, a TDRSS uplink can be scheduled and executed
with an additional few hour latency.   After the observatory receives the new pointing request, slews take place at
0.06\deg/s, so that a full 180\deg\ reorientation takes less than an hour.     So, at best a ToO can take place on a 3 -- 4 hour
timescale, with  12 -- 24 hours being more typical.   

The \nustar\ team monitors supernova alerts as well as other astrophysical alerts daily. 
Other targets monitored for potential ToO triggers include unusual X-ray binary activity, 
magnetar flares, tidal disruption flares, supernova X-ray shocks, and exceptionally energetic GRBs.

\section{Data Analysis}

The \nustar\ Data Analysis Software (NuSTARDAS) consists of a set of general FITS file utility programs and mission-specific data analysis tools 
designed to process the \nustar\ FITS-formatted telemetry data and generate high-level 
scientific data products, including cleaned and calibrated event files, sky images, 
energy spectra, light-curves and exposure maps.

The NuSTARDAS software modules, jointly developed by the ASI Science Data Center 
(ASDC, Frascati, Italy) and the California Institute of Technology (Caltech, Pasadena USA), 
are written in FTOOLS style (the FTOOL protocol was developed by the High Energy Archive Science Research
Center, or HEASARC at NASA GSFC).  The whole package is fully compatible with the HEASoft 
software, maintained and distributed by the HEASARC, where it will be officially integrated. 
The NuSTARDAS input and output files are in FITS format and fully comply with the NASA/GSFC 
Office of Guest Investigator Programs (OGIP) FITS standard conventions. The \nustar\ software 
tasks retrieve the calibration files structured in HEASARC's calibration database (CALDB) 
system.

The software is designed as a collection of modules each dedicated to a single 
function. The input of the NuSTARDAS package is the \nustar\ FITS formatted telemetry data 
(Level 1) produced at the \nustar\ Science Operation Center (\nustar\ SOC) in Caltech. The 
NuSTARDAS data processing is organized in three distinct stages:

\begin{itemize}

\item{Data Calibration:}
Processing of the FITS formatted telemetry to produce Level 1a 
calibrated event files. This step includes the event energy and gain 
 corrections, the search for, and flagging, of hot/bad pixels, the assignment 
of \nustar\ grades, the transformation from raw to sky coordinates, and the 
metrology data processing;

\item{Data Screening:} Filtering of the calibrated event files by applying cleaning criteria on 
specified attitude/orbital/instrument parameters and event properties 
to produce Level 2 cleaned event files;

\item{Product Extraction:} Generation of Level 3 scientific data products (light-curves, energy  
spectra, sky images, Ancillary Response Files, Redistribution Matrix  Files).
\end{itemize}

The package also includes a main script, the \nustar\ pipeline (named `nupipeline'), allowing the user
to run automatically in sequence all the tasks for the data processing. The package has 
also been designed to allow users to reproduce any stage of the data processing that is necessary, for example, because of improved calibration information and/or updated 
software modules, or because the user wishes to use a non-standard data processing procedure. To this end, the `nupipeline' script is the main interface for users.

\section{Education and Public Outreach}

The main purpose of the \nustar\ Education and Public Outreach (E/PO) program is to increase student and public understanding of the science of the
 high-energy Universe. We have developed a multi-faceted program that capitalizes on the synergy of existing high-energy astrophysics E/PO programs
 to support the missionÕs scientific objectives. Our education and public engagement goals are to: facilitate understanding of the nature of collapsed 
objects, develop awareness of the role of supernovae in creating the chemical elements, and to facilitate understanding of the physical properties of the extreme Universe. The E/PO program has additional, more general goals, including increasing the diversity of students in the Science, Technology, Engineering and Mathematics (STEM) pipeline, and increasing public awareness and understanding of the \nustar\ science and technology program.

We are implementing a program that includes elements in higher education, elementary and secondary education, informal education and public engagement. These elements include educator workshops through NASA's Astrophysics Educator Ambassador program, a technology education unit 
for formal educators, articles for Physics Teacher and/or Science Scope magazines, and work with informal educators on a museum exhibit that includes a model of \nustar\ and describes the missionÕs science objectives. Extensive outreach is also underway by members of the science team, who are working with high school students, undergraduates and graduate students. We are also developing printed materials that describe the mission and
special workshops for girls at public libraries in order to improve the STEM pipeline.

\section{The Baseline {\em NuSTAR} Science Program}
\label{sec:science}

The baseline \nustar\ science program began on 1 August 2012, and extends for 25 months, until 30 September 2014.  During this
time, the \nustar\ science team is planning and executing a series of observations aimed at addressing five key scientific goals, described
in detail below, that were the basis for {\em NuSTAR's} selection through NASA's Explorer Program competition.  In addition to these
large projects, the \nustar\ science team has designed and planned a series of observations to be completed during the primary mission
phase.   Tables~5 and 6 summarize the highest priority observations that \nustar\
will undertake during its baseline mission, with the key programs designated as ``Level 1'' targets.   A number of the observations are coordinated
with {\em XMM}, {\em Suzaku}, {\em Chandra}, {\em Swift} or {\em INTEGRAL}.  In addition, for every {\em NuSTAR} field, the {\em Swift} XRT
is taking a 1~ks exposure, in order to constrain the low-energy source properties.   
A Guest Investigator (GI) program will be proposed to NASA, with the aim of extending the mission to allow for broad participation of the community
and an expansion of the scientific scope beyond that of the baseline mission, as well as the additional high priority (Priority A) observations.

\tabletypesize{\scriptsize}
\begin{deluxetable*}{llcl}
\tiny
\tablecaption{Level 1 and Priority A \nustar\ Galactic Targets.}
\tablehead{
\colhead{Working Group} &
\colhead{Target} &
\colhead{Exposure Time} &
\colhead{Notes}}
\startdata
Heliophysics            & the Sun                  & 2 weeks & (2013) \\
Galactic Plane Survey   & Galactic Center survey   & 1.2 Ms  & Level 1 target; 2012 Oct+ \\
                        & Sgr A*                   & 400 ks  & Level 1 target; 2012 Jul+ \\
                        & Limiting Window          & 250 ks  & \\
                        & Sgr B2                   & 200 ks  & (2013) \\
                        & Norma Survey             & 600 ks  & \\
                        & G2 Cloud Infall          & 200 ks  & (2013) \\
                        & Young Massive Clusters   & 300 ks  & \\
SNe+ToOs                & SN Ia to Virgo           & 1 Ms    & ToO \\
                        & CC SN in Local Group     & 1 Ms    & ToO \\
                        & CC SN Shocks             & 200 ks  & ToO \\
                        & Tidal Disruption Flare   & 40 ks   & ToO \\
                        & SN 2010jl                & 46 ks   & 2012 Oct \\
SNRs+PWNe		& SN 1987A		   & 1.2 Ms  & Level 1 target; 2012 Sep+ \\
			& Cassiopeia A		   & 1.2 Ms  & Level 1 target; 2012 Aug+ \\
			& Crab Nebula		   & \nodata & calibration target; 2012 Sep+ \\
			& G21.5-0.9		   & 300 ks  & calibration target; 2012 Jul \\
			& G1.9+03		   & 500 ks  & \\
			& SN 1006		   & 500 ks  & \\
			& MSH 15-52 		   & 150 ks  &  \\
Magnetars		& Magnetar ToO		   & 150 ks  & \\
			& 1E 2259+586		   & 170 ks  & \\
			& 1E 1048-5937		   & 400 ks  & \\
			& AE Aquarii		   & 126 ks  & 2012 Sep \\
			& Geminga		   & 100 ks  & 2012 Sep \\	 
			& PSR J1023+0038	   & 100 ks  & \\
			& 1E 1841-045		   & 45 ks   & 2012 Nov \\
Binaries		& Cen X-4		   & 120 ks  & \\
			& Her X-1		   & 60 ks   & 2012 Sep \\
			& Black hole ToO	   & 200 ks  & \\
			& Accreting pulsar ToO	   & 80 ks   & \\
			& Vela X-1		   & 30 ks   & 2012 Jul+ \\
			& 4U 1820-30		   & 60 ks   & \\
			& V404 Cyg		   & 150 ks  & \\
			& IGR J16318-4848	   & 50 ks   & \\
			& IGR J17544-2619	   & 48 ks   & \\
			& SAX J1808.4-3658	   & 100 ks  & \\
 			& LS~5039	   & 80 ks   &  \\
\enddata
\label{table.galactic}
\tablecomments{Exposure Time refers to baseline planned on-source
integration time. Dates under Notes column refer to when observations
were done.  For longer observations requiring multiple visits, a
plus sign indicates when the observations began.  The priority A target
list is subject to change.}
\end{deluxetable*}

\tabletypesize{\scriptsize}
\begin{deluxetable*}{llcl}
\tablewidth{0pt}
\tablecaption{Level 1 and Priority A \nustar\ Extragalactic Targets.}
\tablehead{
\colhead{Working Group} &
\colhead{Target} &
\colhead{Exposure Time} &
\colhead{Notes}}
\startdata
ULXs            	& NGC 1313                 & 220 ks  & joint \xmm\ \\
                        & IC 342                   & 250 ks  & 2012 Aug, joint \xmm\ \\
                        & NGC 5204                 & 260 ks  & joint \xmm\ \\
                        & M81-X9 (Holmberg IX-XI)  & 200 ks  & 2012 Oct+, joint \xmm\ \\
Extragalactic Surveys   & ECDFS                    & 3.7 Ms  & Level 1 survey; 2012 Sep+ \\
                        & COSMOS                   & 3.7 Ms  & Level 1 survey; 2012 Nov+ \\
                        & Swift/BAT Survey         & 1.6 Ms  & Level 1 survey; 100 targets, 16 ks each+7ks {\em Swift} XRT\\
Blazars+Radio Galaxies  &     Mkn 421       &   300 ks      &   monitoring with Magic, Veritas, Swift   \\
		     & BL Lac     				& 20 ks &  multiwavelength  campaign \\
		     & 3C454.3                              & 300 ks & multiwavelength campaign \\
		     & PKS 2155-304                   & 300 k & multiwavelength campaign \\
		     & B2 1023+25                        & 50 ks  &                                                 \\
AGN Physics             & 3C 273                   & 300 ks  & Calibration target; 2012 Jul \\
                        & MCG 6-30-15              & 180 ks  & joint \xmm\ \\
                        & Ark 120                  & 90 ks   & joint \xmm\ \\
                        & 3C 120                   & 180 ks  &  joint \xmm\ \\
                        & Swift J2127.4+5654       & 180 ks  & 2012 Nov+, joint \xmm\ \\
                        & NGC 4151                 & 150 ks  & 2012 Nov, joint \suzaku\ \\
                        & IC 4329A                 & 185 ks  & 2012 Aug, joint \suzaku\ \\
                        & NGC 3783                 & 300 ks  & \\
                        & MCG 5-23-16              & 300 ks  & \\
Obscured AGN            & NGC 1365                 & 360 ks  & joint \xmm\ \\
                        & NGC 1068                 & 300 ks  & \\
                        & NGC 4945                 & 150 ks  & \\
                        & Circinus                 & 120 ks  & \\
                        & WISE AGN                 & 80 ks   & 4 targets, 20 ks each; 2012 Oct+ \\
                        & BALQSOs                  & 140 ks  & PG~1004+130, PG~1700+518; 2012 Sep+ \\
                        & Mrk 231                  & 50 ks   & 2012 Aug \\
                        & Arp 220                  & 66 ks   & \\
                        & Mrk 273                  & 66 ks   & \\
                        & IRAS 05189$-$2524        & 33 ks   & \\
                        & Super Antennae           & 66 ks   & \\
                        & ULIRG Shallow Survey     & 100 ks  & 5 targets, 20 ks each \\
                        & IC 2560                  & 50 ks   & \\
                        & SDSS Type-2 QS0s         & 40 ks   & 2 targets, 20 ks each; 2012 Oct+ \\
Galaxy Clusters+Relics  & Bullet Cluster           & 250 ks  & 2012 Oct+ \\
                        & Abell 2256               & 250 ks  & \\
Starburst Galaxies      & NGC 253                  & 540 ks  & 2012 Sep+ \\
			& M82			   & 100 ks  & \\
			& NGC 3310		   & 30 ks   & \\
			& Arp 299		   & 180 ks  & \\
			& M83 			   & 180 ks  & \\

\enddata
\label{table.extragalactic}
\tablecomments{Exposure Time refers to baseline planned on-source
integration time. Dates under Notes column refer to when observations
were done.  For longer observations requiring multiple visits, a
plus sign indicates when the observations began.  The priority A target
list is subject to change.}

\end{deluxetable*}

The sections below describe the motivation for the key science programs defined in the
original {\em NuSTAR} proposal to NASA's Explorer Program.

\subsection{Extragalactic Surveys: Probing AGN Activity over Cosmic Time}

\begin{figure}[htbp]
  \centering
  \includegraphics[scale=0.4]{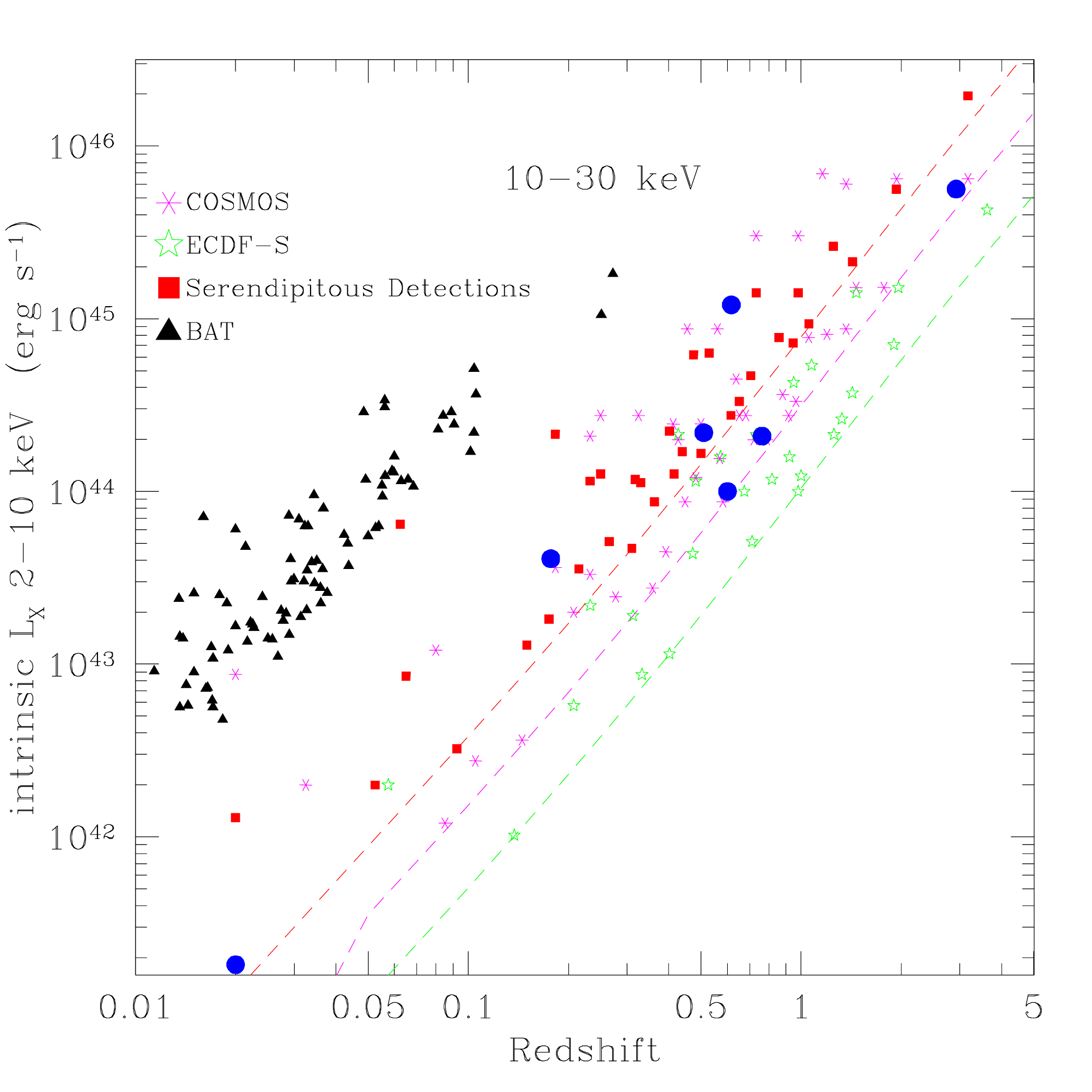}
  \caption{The expected luminosity-redshift distribution for the different \nustar\ Extragalactic surveys.   The
  black diamonds show the distribution of {\em Swift} BAT AGN, the red squares show the expected distribution
  of serendipitous sources in BAT fields, the pink stars show expected results from COSMOS, and the green
  stars from ECDFS.   The large blue circles show the luminosities and redshifts for the first seven serendipitous
  sources discovered in \nustar\ extragalactic fields. }
  \label{fig-lz}
\end{figure}

Resolving the sources that make up the cosmic X-ray background (XRB) has
been a  primary driver for X-ray astronomy over the
last 50~years.  Huge strides have been made 
with sensitive surveys undertaken by the {\it Chandra} and {\it
  XMM-Newton} observatories (e.g.,\ Alexander et~al. 2003; Harrison
et~al. 2003; Luo et~al. 2008; Elvis et~al. 2009; Laird et~al. 2009). These surveys have resolved 70--90\% of
the 0.5--10~keV XRB (e.g.,\ Worsley
et~al. 2005; Hickox \& Markevitch 2006; Xue et~al. 2012), revealing 
obscured and unobscured AGNs out to $z\sim$~5--6
(e.g.,\ Tozzi et~al. 2006; Brusa et~al. 2009; see Brandt \& Alexander
2010 for a recent review). However, {\it Chandra} and {\it XMM-Newton}
are only sensitive to sources emitting below 10~keV, far
from the $\sim$~20--30 keV peak and therefore leaving significant
uncertainties on the properties of the sources that dominate 
(e.g.,\ Ajello et~al. 2008; Ballantyne et~al. 2011). By contrast, the
directly resolved fraction of the XRB at the 20--30 keV peak
from current $>10$~keV surveys is only $\sim$~2\% (e.g., Krivonos
et~al. 2007; Ajello et~al. 2008), and while the sources detected in
{\em Chandra} and {\em XMM-Newton} surveys must resolve at least an
order of magnitude more, no direct observational constraints are
currently available.

The \nustar\ extragalactic survey will provide a significant
breakthrough in revealing the composition of the XRB at the
20--30~keV peak, achieving individual source detection
limits approximately two orders of magnitude fainter than those
previously obtained at $>10$~keV (e.g.,\ Bird et~al. 2010; Tueller et~al. 2010; 
Ajello et~al. 2012).   This will provide a clean
obscuration-independent selection of AGN activity (at least out to $N_{\rm
H}\approx10^{24}$~cm$^{-2}$), required to test claims from {\it Chandra}
and {\it XMM-Newton} surveys that the fraction of obscured AGNs increases
with redshift (e.g.,\ La Franca et~al. 2005; Treister \& Urry 2006). The
\nustar\ extragalactic survey (Table~7) has three primary components:
(i) a deep ($\sim$200--800~ks) small-area ($\sim$~0.3 degree$^2$)
survey of the Extended \chandra\ Deep Field South (ECDFS), (ii) a
medium ($\sim$50-- 100 ks) wider-area (1 -- 2 degree$^2$) survey
of the COSMOS field, and (iii) a shallow ($\sim$15~ks) serendipitous
survey (~$\sim$3 degree$^2$), achieved by targeting 100 \swift-BAT
AGNs.  With this multi-layered approach, the
NuSTAR extragalactic survey will cover a broad range of the flux-solid
angle plane and is expected to directly resolve 30 -- 50\% of the 10 -- 30 keV
background from individual source detections alone (e.g., Ballantyne et
al.  2011).  Sensitive average constraints on
the composition of the XRB will also be obtained by stacking the NuSTAR
data of the faint X-ray sources detected at E$ < 10$~keV by Chandra and
XMM-Newton in the ECDFS and COSMOS fields.

Figure~\ref{fig-lz} shows the expected luminosity-redshift
distribution for the different \nustar\ Extragalactic surveys, derived
using the methodology in Ballantyne {\em et al.} (2011), along with
the positions in this diagram for the first seven
\nustar\ serendipitously detected AGN (Alexander {\em et al.}  2013).
We predict $\sim$~160--200 \nustar-detected sources
($\sim$~65--80 at 10--30~keV) from the medium and deep
\nustar\ survey components, the majority at redshifts between 0.5 and
2 with X-ray luminosities in the range $L_{\rm X} \sim 10^{43} -
10^{45}$~erg~s$^{-1}$. These ranges in redshift and luminosity
correspond to the peak epoch in the black hole accretion history of
the Universe and the knee of the X-ray luminosity function (e.g., Ueda
et al. 2003; La Franca et al. 2005; Aird et al. 2010); thus, \nustar\ should detect the AGN source population that
dominates the cosmic growth of black holes. The shallow serendipitious
survey will achieve two principal aims: (i) provide the identification
of serendipitous sources to trace the bright end of the $>$10~keV XRB
population, and (ii) provide the best global characterization of the
high-energy component of the nearby AGN population (e.g.,\ absorption,
reflection strength, high-energy cut off), to allow for the more
accurate modeling of the X-ray background source populations. We
predict $\sim$~100 \nustar-detected sources in the serendipitous
survey ($\sim$~30 at 10--30~keV), the majority at $z<1$ with $L_{\rm
  X} \sim 10^{43} - 10^{45}$~erg~s$^{-1}$.

\begin{deluxetable*}{lcccc}
\tablewidth{0pt}
\tablecaption{Extragalactic Survey Fields.}
\tablehead{
\colhead{} &
\colhead{} &
\colhead{Exposure} &
\colhead{Depth (10 -- 30 keV)} &
\colhead{Area} \\
\colhead{Survey} &
\colhead{Field} &
\colhead{(ks)} &
\colhead{(erg cm$^{-2}$ s$^{-1}$)} &
\colhead{(deg$^2$)}}
\startdata
Deep    & ECDFS                               & 200--800 & $\sim 2\times10^{-14}$   & 0.3 \\ 
Medium  & COSMOS                              &  50--200 & $\sim 4\times10^{-14}$   & 1 -- 2   \\ 
Shallow & centered on 100 {\it Swift}-BAT AGN & 15       & $\sim 1.5\times10^{-13}$ & 3   \\ 
\enddata
\label{table.egsurvey}
\end{deluxetable*}

\subsection{Galactic Surveys: Studying Compact Objects in the Milky Way}

\nustar's Galactic surveys will extend hard X-ray studies to fainter levels with the aim of detecting large numbers of compact objects 
(black holes, neutron stars, and white dwarfs); determining the hard X-ray morphology of the diffuse emission in the central few dozen 
parsecs of the Galaxy; monitoring the temporal and spectral properties of the supermassive black hole (SMBH) Sgr A*; 
and determining the origin of the Galactic diffuse X-ray background.

The Galactic Center and Galactic Bulge contain 
large numbers of compact objects  (X-ray binaries and Cataclysmic Variables (CVs)) 
of interest for accretion studies, as well as highly-magnetized or rapidly-rotating neutron stars (NS). By detecting 
compact stellar remnants in hard X-rays we can elucidate how populations evolve from groups of single and 
binary high-mass and low-mass stars to reach the ends of their lives as compact objects.  
{\em Chandra} has previously uncovered in excess of 6,000 sources in a $ 2 \times 0.8$~degree region (Muno et al. 2009).  
There is evidence that CVs are the largest component of this population. 
However their nature, whether magnetic or non-magnetic, and their classification, remain uncertain.  Determining 
hard X-ray spectral and timing properties will provide important constraints.

\nustar\ will perform a deep survey of a $ 2 \times 0.8$\deg\ region around the Galactic center and, depending on spectral hardness of the source population, 
should detect up to 100 compact objects.  In addition to the CVs, there are also 30 X-ray sources that have been identified as massive stars 
(Mauerhan et al. 2010), indicating the presence of a significant younger population.  {\em INTEGRAL} observations of this region 
are extremely source-confused, 
and \nustar's vastly better point spread function is required to determine which sources have hard X-ray counterparts.  
It is also unclear from the {\em Chandra} observations whether the massive stars are single or binary and whether any binaries harbor compact objects. 
There are currently only about 100 High Mass X-ray Binaries (HMXBs) known in the entire Galaxy, and \nustar\  (via hard X-ray spectral and timing properties)
can determine whether more HMXBs are present in this region. 

 \nustar\  will also look at other regions of the Galaxy with populations of different ages. 
A major open question in Galactic dynamics is the relative evolutionary states of the different spiral arms. 
It is thought that the Carina arm (Smith and Brooks 2007) is much less evolved than the Norma arm, which, 
in turn, is less evolved than the Galactic center region; however, this has not been demonstrated unambiguously.   
While the number of known HMXBs has increased with {\em INTEGRAL}, it is still likely that the HMXB population is being significantly 
undercounted. If it is true that the Norma region population is significantly younger than the Galactic Center population, 
then one would expect \nustar\ to find a much higher HMXB-to-CV fraction.  

\nustar\ will  survey the $ 2 \times 0.8$~degree  region in Norma covered by {\em Chandra} (Fornasini et al. 2012). 
While {\em Chandra} provides sub-arcsecond source positions and soft X-ray information, a \nustar\ survey of the same region will expand our
understanding of the source populations.  For example, {\em INTEGRAL} results suggest that there may be a large number of low-luminosity HMXBs, an important consideration when the HMXB luminosity function is used as a star formation rate indicator in distant galaxies (Grimm et al. 2003). 
Furthermore, understanding HMXB evolution (as well as the overall number of HMXBs) will help constrain the relative numbers of NS/NS, 
NS/BH, and BH/BH binaries (Tauris and van den Heuvel 2006).
       
The Galactic surveys are also elucidating the morphology of  the central few tens of parsecs around the Galactic Center. 
This region contains X-ray emission from the SMBH Sgr A*, numerous pulsar wind nebulae, supernova remnants 
and molecular clouds (Baganoff et al. 2003).  Of particular interest for \nustar\ are the flares from Sgr A*. 
These flares, seen about once per day, have been detected from radio wavelengths all the way to the soft X-ray band, but only
hard X-ray upper limits existed until \nustar\ observed a large flare in July 2012.
The \nustar\ data, in conjunction with observations at lower energies, put tight constraints on emission mechanisms (Barri{\`e}re {\em et al.} 2013).
Further campaigns are planned for next year, to be coordinated with {\em Chandra}, {\em XMM} and Keck.
{\em INTEGRAL} has observed this region extensively (Belanger et al. 2006); however, it is severely
source-confused given the 12\arcmin\ FWHM PSF.  With sub-arcminute angular resolution and $\sim 3$\arcsec\ localizations, 
\nustar\ is searching for hard X-ray sources and mapping the complex diffuse emission surrounding Sgr A* out to
$\sim50$~keV. 

\nustar\ will also map diffuse hard X-ray emission from molecular clouds. 
In particular Sgr B2 has recently attracted much attention, since its diffuse Fe-K and continuum X-ray emission are 
time-correlated,  as would be expected if the emission is in response to intense flaring activity of some bright source, possibly Sgr A*, 
in the past (Sunyaev and Churazov 1998).  
A less likely alternative is injection of cosmic-rays into the cloud (Yusef-Zadeh et al. 2007) with subsequent X-ray emission. 
These processes predict distinct morphological differences in the hard X-ray continuum emission, enabling us to distinguish between the two models; furthermore, by measuring the strength of the inverse Compton continuum emission, we can constrain the density
of the cloud, and therefore the required input energy (Ponti {\em et al.} 2012).

The combination of the $2 \times 0.8$~degree Sgr A* survey, and an observation of the low column density Limiting Window, 
\nustar\ should also be able to resolve the nature of the Galactic diffuse X-ray background (XRB).   Likely due to unresolved
point sources (Revnivtsev {\em et al}. (2009), Hong (2012)), it is not clear what fraction is due to magnetic CVs vs coronally active
giants, which would have relatively soft spectra.  \nustar's ability to measure hard X-ray spectral properties should settle this
issue.

\subsection{Broadband Observations of Blazars and the Nature of Relativistic Jets}

Many active galaxies possess prominent relativistic jets, and if they
point close to our line of sight, jet emission can dominate 
the radiative output of the object from radio to very energetic $\gamma$-rays.  
Termed blazars, the emission from these sources is highly variable, by up to factors of 100 in the optical
(see, e.g., Wagner and Witzel 1995) and 1000 
in $\gamma$-ray (see, e.g., Ackermann et al. 2010 and Wehrle et al. 2012). Blazar spectral energy distributions 
generally consist of two prominent, broad humps, one peaking in the 
mm-to-X-ray spectral range, and the other in the hard X-ray - to - high energy 
$\gamma$-ray range.  Synchrotron emission is responsible for the low-energy
peak, while inverse Compton produces the gamma-ray hump.  General 
coincidence of variability patterns in the far IR - to optical 
band with the $\gamma$-ray band suggests that the two processes are likely 
to originate from the same population of highly relativistic electrons, 
with Lorentz factors $\gamma_{\rm el}$ reaching $10^{6}$ or higher.  
The situation in the X-ray band is less clear, since the variability 
is only modestly correlated with that in the optical and $\gamma$-ray bands 
(see, e.g., Bonning et al. 2009, Hayashida et al. 2012).

Hard X-ray measurements are  important 
diagnostics of the content of the jet, as well as of the processes 
responsible for the acceleration of particles to the very highest energies.  
In luminous blazars - often associated with powerful quasars - 
the X-ray spectra are generally hard, suggesting that the X-ray 
band represents the onset of the inverse Compton component (see, e.g., 
Kubo et al. 1998) .  The low-energy end of the Compton
component is produced by particles with the highest energies, since
the particle spectrum falls steeply with energy so that
X-ray observations constrain 
the total particle content of the jet (see, e.g., Sikora \& Madejski 
2000).  Conversely, in the low-luminosity objects, the X-ray band 
is thought to be the high-energy end of the synchrotron component, and is therefore
produced by the most energetic ``tail'' of the electron spectrum 
(cf. Takahashi et al. 1996).  Hard
X-ray observations  conducted jointly with the TeV $\gamma$-ray 
observations probe electrons with most extreme energies.  This is especially 
important for the most energetic TeV emitters, which reach the highest 
apparent  values of $\gamma_{\rm el}$, since they become radiatively 
inefficient in the Klein-Nishina regime. 
Measuring the most energetic synchrotron emission allows one to distinguish between  intrinsic and extrinsic cutoff mechanisms, 
and sets clear constraints on 
$\gamma_{\rm el}$.   \nustar's large bandwidth provides a unique tool for 
such investigations. 

The extreme variability of blazars provides an opportunity 
to study the evolution of the radiating particles' energy distribution, from the 
acceleration stage through their luminous emission.   Broad-band
studies can probe the 
conversion of jet bulk kinetic energy 
to ultra-relativistic electrons and subsequently to photons.  The relative temporal 
variability in different spectral bands, from radio through TeV, provides 
an important handle on the location of the energy dissipation 
as a function of distance from the central black hole (Sikora et al. 2009).   

As the main component of its blazar program,
\nustar\ will participate in large multi-wavelength campaigns of two luminous 
blazars showing quasar-like behavior, and two low-luminosity sources in
an active state.  Extreme activity is unpredictable (see, e.g., Wagner and 
Witzel 1995;  Abdo et al. 2010b), so that 
targets must be chosen by monitoring activity in the {\em Fermi} band.
In the quasar class, 
3C279 ($z = 0.524$) and 3C454.3 ($z = 0.859$) are the most promising
targets.  Both have shown extreme 
variability in Fermi on a time scale of $\sim 3$ days 
(Abdo et  al. 2010b).  \nustar\ will observe with a daily cadence
of 10~ks pointings  with facilities from radio to $\gamma$-ray with
sampling as simultaneously as possible.

The best candidates for low-luminosity blazars are Mkn 421 and PKS 2155-304.
The most extreme variability for these is seen in the TeV band, 
on time scales of an hour or less (see, e.g., Benbow et al. 
2008,   Galante \etal\ 2011).   Assuming these stay active, \nustar\
will observe simultaneously with MAGIC and Veritas (for Mkn 421),
and H.E.S.S. (for PKS 2155-304) for several nights each month during
the TeV telescopes' observing seasons.  The continuous pointings will
be  strictly simultaneous (modulo Earth occultations).  Preliminary data 
obtained for Mkn 421 during the \nustar\ calibration phase already indicate 
that such measurements of time-resolved spectra with \nustar\ are feasible, 
where a $\sim 15$ ks observation demonstrably samples the spectrum to 50 keV.  

\nustar\ will also make shorter observations of several blazars with particularly unusual properties.  
Candidates include 1ES 0229+200, where hard X-ray and TeV $\gamma$-ray emission have 
interesting implications for the intergalactic magnetic field 
(see Kaufmann \etal\ 2011 and 
Vovk \etal\ 2012 ), and AO 0235+164 (see Ackermann \etal\ 2012 and Agudo \etal\
2011).  The scheduling of these targets will depend on them entering
an active state as determined by {\em Fermi}.

\subsection{Non-thermal Radiation in Young Supernova Remnants}

Young ($\tau \simless 1000$~year-old) supernova remnants are priority
targets for \nustar, which will provide the first sensitive imaging
observations of non-thermal emission above 10~keV.  X-ray synchrotron
emission is known to extend to high energies in several young remnants
(see Reynolds 2008 and Vink 2012 for reviews) and possibly in others;
its characterization addresses questions of particle acceleration and
the origin of cosmic rays (e.g., Reynolds 1998; Warren \etal~2005).
Remnants younger than 1000~years may also show evidence of relatively
long-lived radioactive nuclides, primarily $^{44}$Ti, which is of
importance in understanding explosion mechanisms (Arnett et al.~1989;
Timmes et al.~1996; Magkotsios \etal.~2010).

The primary remnants for these studies are Cas~A, SN~1987A, and
G$1.9+03$ -- objects where $^{44}$Ti emission has been detected, and
which are also interesting for studying the hard synchrotron
continuum.  $^{44}$Ti (mean life 85 yr) decays by electron capture to
$^{44}$Sc, leaving an inner-shell vacancy which can be filled by a
transition producing a 4.1~keV X-ray.  Nuclear de-excitation lines at
68 and 78~keV and 1.157~MeV are also produced. So far, the gamma-ray
lines have been seen only from Cas~A, by Beppo/SAX (Vink et al. 2001)
and IBIS/ISGRI on {\em INTEGRAL} (Renaud et al. 2006), and recently from
SN1987A by IBIS/ISGRI~(Grebenev et al.~2012).  The $^{44}$Sc X-ray
line has been seen only in the youngest Galactic remnant G1.9+0.3
(Borkowski et al. 2010), whose age is about 100 yr (Carlton et
al.~2011).

Theoretical predictions for the initial synthesized mass of $^{44}$Ti
in core-collapse SNe range from $(0.1 - 3) \times 10^{-4}$~M$_\odot$
(Timmes et al.~1996; Thielemann, Hashimoto, \& Nomoto 1991; Woosley \&
Hoffman 1990, Young {\em et al.} 2006).  In Cas~A the inferred yield is about $2 \times
10^{-4}$~M$_\odot$ of $^{44}$Ti, at the upper end of the range of
expectations.  The yield in SN1987A is also surprisingly
high; $3.1 \pm 0.8 \times 10^{-4}$~M$_\odot$ (Grebenev et al. 2012).  
\nustar\ is executing several long observations of Cas~A,
with over 1.2~Ms of total exposure will make the first spatial maps
of the remnant in $^{44}$Ti.  The asymmetry of the emitting region will provide
strong constraints on the asymmetry of the explosion.   
Similarly in 1987A, assuming the {\it INTEGRAL} detection is robust, \nustar\'s 1.2 Ms integration will significantly
improve the yield measurement, and will constrain the line
width as well as search for hard X-ray emission from any compact
object.

The very high expansion velocities detected spectroscopically from
G1.9+0.3 ($>$12,000~km~s$^{-1}$), a century after the explosion, argue
fairly strongly for a Type~Ia origin, supporting arguments derived from the remnant
morphology and the absence of a PWN (Reynolds et al.~2008). However,
the SN-type question is still open.  The $^{44}$Sc 4.1 keV line intensity
detected from G1.9+0.3 (Borkowski et al.~2010) implies about
$10^{-5}$~M$_{\odot}$ of $^{44}$Ti (for an assumed age of 100 yr).
The predicted fluxes in the 68 and 78 keV lines are about $4 \times
10^{-6}$~ph~cm$^{-2}$~s$^{-1}$ each. A detection of
$^{44}$Ti from a firmly classified Type~Ia remnant will be a very important
constraint on explosion models. If G1.9+0.3 turns out to be a
core-collapse remnant, having another example alongside Cas A and SN
1987A will be important in isolating the important variables
determining $^{44}$Ti production. 

All Galactic SNRs less than 2000 yr old, and some older examples, show evidence
of synchrotron X-ray emission (Reynolds 2008; Allen, Gotthelf, \&
Petre 1999).  In all observed cases, the intensity is below the
extrapolation of the radio synchrotron spectrum (Reynolds \& Keohane
1999), indicating that some processes (radiative losses, finite
remnant age or size, or particle escape) are beginning to cut off the
electron acceleration process at high energies.  Considerably more
information on particle acceleration is encoded in the detailed shape
of the spectral steepening than in the power-law portion of the
spectrum at lower energies.  Most inferences to date are based on
observations below about 8~keV, where power laws are indistinguishable from more detailed
models.  Both Cas A and G1.9+0.3 have strong X-ray synchrotron
components; in the case of Cas A, spatially integrated observations to
80 keV have been performed (Favata et al.~1997; Allen et al.~1999;
Maeda et al.~2009). \nustar\ will provide spatially resolved maps.   
The integrated spectrum of G1.9+0.3 is dominated by
non-thermal emission below about 7 keV (Reynolds et al.~2008), but no
observations have been done at higher energies.  Since G1.9+0.3 is the
only Galactic SNR currently growing brighter (Carlton et al~2011), the
behavior of its spectrum at higher energies is of great interest.  For
SN 1987A, X-ray emission below 10 keV is composed of two thermal
components (e.g., Sturm et al.~2010), but our long observation to
search for 68 keV line emission should be able to detect or constrain
any non-thermal continuum at higher energies.

\subsection{Dynamics in Core Collapse and Ia Explosions: Target of Opportunity Observations of Supernovae}

\nustar\ will provide new insights into supernova explosions in our local Universe by studying the hard X-ray emission from SN Ia explosions out to the Virgo cluster ($\sim$14 Mpc) and  core collapse supernovae in the Local Group ($\sim$4 Mpc). This emission arises from the Compton down-scatter of $^{56}$Ni decay gamma-rays, and both the lightcurve and spectrum emerging from the 
supernova at hard X-rays is sensitive to the mass and abundance distributions in the explosion. 
Detailed simulations of SN~Ia have shown that this hard X-ray emission is a strong diagnostic of the surface 
composition and flame propagation, and larger $^{56}$Ni mass produces higher hard X-ray flux levels that peak at 
earlier times (Maeda {\it et al.} 2012).  Typical peaking times are expected to be between 10-30 days for SN Ia, and between 
150-250 days for core collapse supernovae. 
Our primary criteria for triggering a supernova TOO is to have a spectroscopically confirmed supernova identified before peak within the \nustar\ 
sensitivity distances.

\subsection{Additional Priority Science Observations}

\nustar\  will pursue a range of science objectives significantly broader than the key projects described above.
The sections below detail additional science objectives, and describe which targets 
will be observed during the baseline mission.   Fully exploiting the range of opportunities will require 
a Guest Investigator program, to be proposed to NASA as an extension to the baseline mission.
Approximately 70\% of the total observing time in the baseline mission is allocated to the Priority A targets plus the key (Level 1)
programs; the remaining time
(30\%) is kept in reserve for followup of unanticipated discoveries, as well as additional ToO observations.   

\subsubsection{Ultraluminous X-ray Sources}

Ultraluminous X-ray Source (ULX)  luminosities (0.5--10~keV) exceed $\sim 3 \times 10^{39}$ erg/s
(extending up to $10^{42}$ erg/s; Farrell et al. 2009), significantly
above the isotropic Eddington luminosity for stellar mass accretors. 
Hypotheses are that ULXs contain stellar mass BHs in an extreme
version of the very high state of Galactic BHs -- the `ultra-luminous'
or wind-dominated state (Gladstone et al. 2009).  In this state,
sources can radiate above the Eddington limit and exhibit a soft
spectrum cutting off just below 10 keV.
Alternatively, the emission may display some anisotropy due to
geometrically thick disks that act as a funnel for X-ray photons produced in the
inner disk region (King 2009). Most intriguing is the idea that ULXs
- in particular the very luminous ones - harbor intermediate mass BHs
($10^2$--$10^5$ $M_{\odot}$; IMBH) accreting at sub-Eddington rates.
The existence of IMBHs is vitally important to establish, because they
may provide seeds that form supermassive BHs via mergers or accretion
episodes (e.g., Micic et al. 2007).

\nustar\ is observing a sample of bright ULXs simultaneous with \xmm\  as part of
approved AO-11 and 12 programs.   Working together, \xmm\ and \nustar\ can provide powerful observations
aimed at elucidating the nature of ULXs.  \xmm\ covers the spectral
region where the accretion disk shines (peaking at $\sim1$~keV), and
\nustar\ will explore whether Comptonization of soft photons on hot
electrons occurs, and will enable searches for signatures of
relativistic beaming.  In Galactic binaries, contemporaneous broad-band spectroscopy and relative hard/soft band variability has been a
powerful tool for understanding accretion physics and the nature of
the compact object.  It will likewise be the case for ULXs.  In addition, \nustar\
is observing NGC~1365 and M82 as part of the obscured AGN and starburst
galaxy programs.

\subsubsection{Accretion Physics in Active Galactic Nuclei}

\nustar\ will make numerous contributions to understanding the
physical mechanisms at work in radio-quiet AGN by addressing three long-standing and hotly debated issues:

{\bf Is relativistic reflection the right model?}   
In radio-quiet AGN the primary X-ray emission is reflected off  the
accretion disk.   This reprocessing results in a  `Compton hump'  - a continuum excess peaking
in the \nustar\ band at around 30~keV  (George \& Fabian 1991, Matt et al. 1991), as well as
several fluorescent lines, the most prominent being the iron 
K$\alpha$ (Reynolds et al. 1995).  Matter velocities are high, and general and special relativistic effects 
are very strong in the vicinity of the BH, modifying emission line profiles
(Fabian et al. 1989; see Fabian et al. 2000, Miller 2007 for reviews). 
This results in two asymmetric peaks (the bluest being the brightest) and an extended red wing.

{\it ASCA}'s discovery  of a prominent red tail in the iron line profile
of the Seyfert 1 galaxy MCG-6-30-15 (Tanaka et al 1995) - later confirmed
in this and other objects 
(e.g., Nandra et al. 2007; de la Calle Perez et al. 2010; Brenneman et al. 2011) -
seem to validate this model.  However, this red wing may not
be intrinsic, but due to the presence of ionized absorbers partially
covering the primary X-ray source (Turner \& Miller 2009).  Clear-cut 
observational proof is still missing.   \nustar\  is providing definitive observations 
validating the reflection model (Risaliti {\em et al} 2013). 
While the two models are largely degenerate below 10 keV, they make very different
predictions in hard X-rays. 

{\bf The spin of the black hole.} Assuming that the relativistic reflection model is the correct one
\nustar,  in coordination with \xmm\ and \suzaku, can
determine the spin of the black hole, a key parameter in understanding
the evolution of AGN and the history of black hole growth
(Berti \& Volonteri 2008, 
Volonteri et al. 2012). The black hole spin can be estimated as a consequence of its
affect on the innermost disk radius ({\it e.g.},  Laor 1991), 
which  can be measured by fitting the iron line profile, and in particular
the low-energy extension of the red wing.
While \xmm\ and \suzaku\ observations
alone  already provide precise measurements from the statistical point of 
view (e.g., Brenneman et al. 2011), but systematic uncertainties are still 
problematic.    \nustar\ has already begun a program of observations with \xmm\ and \suzaku\
that will provide much tighter constraints on the emission continuum, 
allowing significantly more precise spin measurements.

{\bf The temperature of the hot corona.}  There is consensus that the primary X-ray emission
is due to Comptonization of UV/soft X-ray accretion disk photons in a hot corona, 
the properties of the corona, and the values of its two main parameters, the 
temperature and the optical depth, are largely unknown.  
This ignorance hampers our understanding of fundamental properties
such as the relation between the
corona, accretion disk, and the energy budget of AGN.
Previous studies (e.g., Perola et al. 2002) indicate coronal temperatures 
in the 50-200 keV range, as derived from the measurement of high-energy 
cut-offs observed in the power-law continua of several AGN.  However,
the quality of the measurements obtained with non-focusing 
hard X-ray telescopes ({\it BeppoSAX}, {\it INTEGRAL} and \suzaku) 
is too poor to ensure that the spectral complexity was 
dealt with properly and the cut-off energy correctly estimated. 

\nustar\ observations will allow us to reliably determine the properties of
the hot emitting corona for the first time. Simulations indicate that 
the coronal temperatures can be estimated with errors of only 10\% 
for the brightest AGN if the temperatures are not larger than 100 keV,
and with errors of 20-30\% for temperatures as large as 200 keV.

\subsubsection{Galaxy Clusters and Radio Relics}

Due to the limitations in sensitivity and angular resolution of previous hard X-ray missions, there are few
constraining observations of galaxy clusters above 10~keV.  Many merging clusters contain hot
gas with temperatures that can only be poorly constrained by data below 10~keV.  \nustar's 
unprecedented sensitivity in the 10-20 keV energy band will allow us to
provide significant constraints on the hot thermal emission expected, and sometimes observed,
in merging systems.  The first \nustar\ observation of a galaxy cluster has recently been conducted on the Bullet Cluster; this well known system hosts hot thermal emission 
(Markevitch \& Vikhlinin 2007).

In addition, \nustar\ can map non-thermal emission associated with populations of relativistic electrons
in clusters from which radio synchrotron emission has been detected.   These detections have
a rather controversial history (Wik {\it et al.} 2012), with measurements of ``hard tails'' 
having been made by some authors and challenged by others.  If the claims
are true, the high sensitivity and low background of the \nustar\ mission will afford
robust, high-significance detections.  If, conversely, the claims are false
\nustar\ observations will provide tighter upper limits and, possibly,
some detections.

\subsubsection{Nearby Star Forming Galaxies}

The \nustar\ starburst program has two principal components: 1) an
intensive program on NGC 253 ($d=2.6$ Mpc) for 500~ks in three 165~ks 
observations with coordinated {\it Chandra}, VLBA and (automatic) 1--2 ks \swift-XRT visits (completed  Nov 2012),  
and 2) a survey of five additional nearby (4--40~Mpc) galaxies with
30--180~ks exposures to build up a statistical sample.
Unless all X-ray binaries in these galaxies are unexpectedly very soft, we
expect to detect all of the galaxies out to energies of $\sim$20~keV.

Given the relatively high specific star-formation rate of NGC~253, its
X-ray binary population is expected to be primarily high-mass X-ray
binaries (HMXBs) in contrast to the low-mass X-ray binaries which are the
dominant population in the Milky Way. Our monitoring consisted of
three observational periods which included
20~ks with {\em Chandra}, and 28~ks with the VLBA. The monitoring was
designed to (1) sensitively isolate the locations of X-ray binaries,
(2) determine the nature of the accreting compact objects via their
0.5 -- 30 keV spectral properties, and (3) identify interesting
flaring X-ray/radio sources as they make spectral state
transitions due to variability in their accretion. 

As part of the baseline program, \nustar\ will survey five additional normal/starburst galaxies:
M82, M83, NGC 3256, NGC 3310 and Arp 299. The \nustar\ observation of M82 will likely be dominated by the
``famous" ULX M82 X-1 (Kaaret et al. 2006).  The
exposure times for these five galaxies ranges from 30--180~ks,
driven by the goal of producing a 5$\sigma$ 10--20~keV detection
under conservative assumptions (moderately soft spectral shape).
It is expected that these observations will result
in strong detections of at least the brightest several
sources in each galaxy. The main goals of the survey program are 1) to characterize the typical starburst spectrum above 10 keV, 2) to identify the nature of individually-detected X-ray sources (neutron star vs. black hole candidate), and 3) to look for short-term (hours to weeks) variability and establish a baseline for long-term variability studies (weeks to years)

\subsubsection{Magnetars and Rotation Powered Pulsars}

\nustar\ will observe a sample of
rotation-powered pulsars (RPPs) and magnetars, weighted toward the magnetars given the surprising high-energy turnover in their X-ray spectra discovered by {\it INTEGRAL}
and {\it RXTE} (Kuiper et al. 2006).  This emission
is both unpredicted and not understood, though some models have been
suggested (e.g., Heyl \& Hernquist 2005; Baring \& Harding 2007;
Beloborodov \& Thompson 2007; Beloborodov 2012)
and some intriguing correlations noted (Kaspi \& Boydstun 2010;
Enoto et al. 2011).  Specifically, among the Priority A
magnetar targets is 1E 2259+586, only marginally detected in the hard
X-ray band by {\it RXTE} and {INTEGRAL}.
\nustar\ will provide the first high quality broad-band X-ray spectrum
for this target, which has the most extreme spectral turnover yet seen in
any magnetar. In addition,  will observe 1E 1048$-$5937, the lone
`classical', regularly monitored magnetar for which no hard X-ray emission
has yet been detectable.  \nustar\  will also observe
1E 1841$-$045, the magnetar located at the center of the supernova remnant
Kes 73 which is among the first magnetars detected as a hard X-ray emitter (Kuiper et al. 2006).  
With a relatively short exposure, \nustar\
can measure an exquisitely precise spectrum, clearly defining the turnover
region as well as the pulsed fraction as a function of energy.  Finally,
\nustar\ plans a target-of-opportunity observation of a magnetar in outburst
and its subsequent relaxation, with the goal of monitoring simultaneously the behavior of the soft and hard X-ray components to see whether they vary in concert.

For RPPs, \nustar\ will observe the Geminga pulsar,
among the very brightest gamma-ray sources known, which shows relatively faint, mainly thermal X-ray
emission (e.g. Kargaltsev et al. 2005) and is an archetype for
X-ray faint/gamma-ray bright RPPs.  In addition,  \nustar\
will observe the millisecond pulsar PSR J1023+0038.  This pulsar is
a unique transition object between the  millisecond pulsar and low-mass X-ray
binary evolutionary phases.  X-ray emission observed with
{\it XMM} and {\it Chandra} (Archibald et al. 2010; Bogdanov
et al. 2011) shows a hard power-law tail suggesting
detectability well above 10 keV with {\it NuSTAR}.

\subsubsection{Pulsar Wind Nebulae}

Pulsar-wind nebulae exhibit radiation from extremely energetic
particles produced by pulsars and processed through a relativistic
wind termination shock.  The nature of magnetospheric acceleration,
of the behavior of the cold relativistic wind interior to the termination shock, and the
particle-acceleration properties of the relativistic shock are all important open issues. Even the composition
of the wind is unknown, although it represents a major part of the pulsar's energy loss.

As part of the in-flight calibration program, \nustar\ will observe
the most famous PWN, the Crab Nebula, as well as G21.5-0.9.  For the
Crab Nebula, the integrated spectrum is well-known through the NuSTAR
bandpass (though there may remain some disagreements among instruments;
Kuiper et al. 2001, Jourdain \& Roques 2009). However, no high-quality images
of the nebula have been made above 10 keV.   
A slow decrease in the effective size of the
nebula with X-ray energy is predicted  by the model of Kennel and Coroniti (1984). 
\nustar\ will test this prediction with much greater
sophistication than possible with scanning collimator and occultation measurements.  
High-resolution {\sl Chandra} and {\sl XMM-Newton}
observations of the Crab Nebula and other young PWNe typically show a
``jet-torus'' structure on arcminute or smaller scales (see Gaensler
\& Slane 2006 for a review).  Below 10 keV, the Crab Nebula torus has
a harder spectrum than the rest of the nebula (Mori et al.~2004), so
changes in morphology as a function of X-ray energy above 10 keV are
expected. Similarly, in G21.5-0.9 \nustar\ can determine the
integrated spectrum to high accuracy, measure the evolution of the PWN
size as a function of energy, and search for X-ray pulsations from the
central neutron star.   A third PWN, SNR MSH 15--5{\sl 2} will also
be observed as part of the baseline science program.

\subsubsection{Obscured AGN}

The existence of Compton-thick AGN (CTAGN), with their telltale
10 -- 13~keV reflection hump, has been invoked regularly
as a major component to explain the steep shape and 30~keV peak of
the cosmic X-ray background (e.g., Comastri {\em et al} 1995; Gilli {\em et al.} 2007).
Despite their importance, however, prior to the launch
of \nustar, only a handful of CTAGN were directly detected above
10~keV, most of which are only weakly Compton-thick.  While this
was sufficient to confirm their existence, physical constraints
generally are remarkably poor, and the complex range of expected
spectral features (Murphy \& Yaqoob 2009) remain largely untested.  Furthermore,
only half the such known sources belong to flux-limited surveys, all
at $z \leq 0.015$ and thus have very modest intrinsic luminosities.

\nustar\ plans a comprehensive program studying CTAGN.  One aspect
of this program is deep observations of famous, bright sources such
as NGC~1068 and NGC~1365 which span a range of obscurations.  These
data will provide detailed high energy spectra,
including measurements of the reflection humps, covering fractions,
and variability.  Shallower observations of fainter sources will
investigate the range of obscured AGN properties.  The latter
category includes well-known ultraluminous infrared galaxies (Risaliti {\em et al.} 2000, Teng {\em et al.} 2005), 
a few optically bright
broad-absorption lined quasars (BALQSOs) with surprisingly faint
soft X-ray fluxes (Gallagher \etal\ 1999,  Saez {\em et al.} 2012),
optically selected type-2 quasars identified from the Sloan Digital
Sky Survey (Zakamska \etal\ 2004), and heavily obscured AGN
candidates recently identified by the {\it Wide-field Infrared
Survey Explorer} (Eisenhardt \etal\ 2012).

\subsubsection{Galactic Binaries}

Hard X-ray observations of accreting black holes 
provide a probe of the inner regions of the accretion disk where strong 
gravity prevails (Remillard \& McClintock 2006; Psaltis 2008).  Neutron stars are exotic 
objects where densities, magnetic fields, and gravitational fields are 
all extreme \cite{lp04,lp07,tru78,coburn02}.  Studies of X-ray binaries 
are important for understanding accretion disks and determining how 
relativistic jets, which are seen from X-ray binaries as well as
from supermassive black holes \cite{fender06}, are produced.  Constraining
the properties of the compact objects themselves has implications
for gravitational and nuclear physics.

For faint X-ray binaries \nustar\ will provide the first hard X-ray studies of black hole and neutron star 
X-ray transients at their lowest mass accretion rates, 
providing a new window on accretion processes and emission mechanisms 
in quiescence.  {\em NuSTAR} will investigate the origin of the hard 
X-ray emission from the neutron star transient Centaurus~X-4
\cite{rutledge01,cackett10} and from the black hole transient V404~Cyg
(Narayan {\em et al.} 2007, Bradley {\em et al.} 2007, Corbel {\em et al.} 2008).

{\em NuSTAR}'s spectroscopic capabilities will be used for studies
of bright X-ray binaries.  Cyclotron resonance scattering features
appear as ``cyclotron lines'' at energies of 10--100\,keV, providing
the only direct measurement of an accreting neutron star's magnetic 
field strength.  The most advanced theoretical model for these 
lines \cite{schoe07} will be applied to {\em NuSTAR} spectra of 
High-Mass X-ray Binaries (HMXBs) such as Hercules~X-1, Vela~X-1, and 
accreting pulsar transients; we will also search for cyclotron lines 
from members of newly discovered classes of HMXBs 
\cite{walter06,negueruela06}.  In addition, {\em NuSTAR} will target 
accreting black hole systems, such as Cygnus~X-1 and transient black 
hole systems, in order to study the Compton reflection component 
\cite{mz95}.  Both the strength of this component and the level of 
relativistic distortion in the emission lines and absorption edges 
provide constraints on the geometry of the inner accretion disk and 
on the black hole spin \cite{miller07}.  Another {\em NuSTAR}
project will be to study neutron stars that exhibit X-ray bursts
that occur when accreted material undergoes a thermonuclear runaway
(Lewin {\em et al} 1993).  {\em NuSTAR} will observe 4U~1820--30 when it is 
actively producing X-ray bursts to search for absorption edges that 
have been tentatively detected with lower resolution hard X-ray 
detectors \cite{iw10}.  Such edges have the potential to provide 
detailed information about the nuclear burning sequence, burst ignition, 
and the conditions near the neutron star surface.

{\em NuSTAR} observations are planned with timing as the primary 
goal.  SAX~J1808.4--3658 is a transient accreting millisecond pulsar that 
exhibits 400\,Hz pulsations during outbursts (Wijnands \& van der Klis).  However, as
the mass accretion rate drops, it is predicted that the ``propeller
effect'' (Illarionov \& Sunyaev 1975) will eventually cause the accretion onto the 
magnetic poles to cease due to the centrifugal barrier, and {\em NuSTAR} 
will test this by observing SAX~J1808.4--3658 at the end of an outburst.
Timing will also be important for carrying out phase-resolved spectroscopy
of the cyclotron line studies.  Finally, {\em NuSTAR} plans to target  the gamma-ray
binary LS 5039 to search for pulsations and to measure how the hard X-ray spectrum
changes with orbital phase.

\subsubsection{Solar Physics}

The \nustar\ spacecraft and instrument are capable of pointing at the
Sun, an unusual ability in a space telescope designed for
astrophysical observations.    \nustar\ will be able to
detect hard X-rays from solar activity at levels two hundred times
fainter than permitted by the background-limited sensitivity of the {\em Reuven Ramaty
High Energy Solar Spectroscopic Imager (RHESSI)}, another NASA Small
Explorer and the state-of-the-art for a solar hard X-ray mission.  The  solar program during the primary
mission is planned as three weeks of pointings,  both to active regions
and quiet parts of the solar disk and limb, with two principal science goals: the search for hard X-rays from very small
events (nanoflares) and the study of faint hard X-ray sources high in
the solar corona associated with both large and small flares.  These
observations will take place near the end of the \nustar\ primary mission.

\section{Summary}

The \nustar\ observatory, launched in June 2012, is operating nominally on-orbit, with performance fully consistent
with pre-launch predictions.   Completion of the initial calibration indicates that the cross normalization and relative response
agree well with other soft X-ray telescopes below 10~keV.  
The science program for the two-year baseline science mission includes both
key projects as well as a diverse set of investigations designed to provide a broad scientific scope.  The mission 
life is limited by the orbit altitude and single-string design; however with the $>600$~km near circular orbit the
\nustar\ could continue operations for up to ten years.     After completion of the baseline program, the \nustar\ team
will propose an extended mission that emphasizes a guest investigator program, along with continued 
Extragalactic and Galactic surveys.  This will provide opportunities to expand the multi-faceted program through dedicated
\nustar\ pointings, to follow-up on discoveries, and for further joint programs with XMM and {\em Chandra},  as well
as a broader Target of Opportunity program.

\acknowledgments
\nustar\ was developed and is primarily supported by the National Aeronautics and Space Administration
(NASA) under NASA No. NNG08FD60C.   Support for development was provided by the National Space Institute,
Technical University of Denmark.   The Malindi ground station is provided by the Italian Space Agency (ASI),
and support for science software development by the ASI Science Data Center (ASDC).   Science team members
acknowledge support from Centre National d'Etudes Spatiales (CNES) (D Barret, M. Bachetti), the NASA Postdoctoral
Program (D. Wik), Leverhulme Research Fellowship and Science and Technology Facilities Council (DMA), NSF AST (DRB),
U.S. DOE/LLNL (WWC,MP, JV), NSERC, CIFAR, FQRNT, Killam Research Fellowship (VK).

\clearpage

\bibliographystyle{apj1b}   

\clearpage

\end{document}